%% file: main.tex
\documentclass[conference]{IEEEtran}
\IEEEoverridecommandlockouts
\usepackage{cite}
\usepackage{amsmath,amssymb,amsfonts}
\usepackage{algorithmic}
\usepackage{graphicx}
\usepackage{textcomp}
\usepackage{xcolor}
\usepackage{subcaption}

\def\BibTeX{{\rm B\kern-.05em{\sc i\kern-.025em b}\kern-.08em
    T\kern-.1667em\lower.7ex\hbox{E}\kern-.125emX}}

\makeatletter
\newcommand{\linebreakand}{%
  \end{@IEEEauthorhalign}
  \hfill\mbox{}\par
  \mbox{}\hfill\begin{@IEEEauthorhalign}
}
\makeatother

\newcommand\xingfu[1]{{\color{black}#1}}

\begin{document}
\title{ytopt: Autotuning Scientific Applications for Energy Efficiency at Large Scales}

%\if 0

\author{\IEEEauthorblockN{Xingfu Wu, Prasanna Balaprakash, Michael Kruse, Jaehoon Koo, Brice Videau, Paul Hovland, Valerie Taylor}
\IEEEauthorblockA{Argonne National Laboratory, Lemont, IL 60439 \\
Email: \{xingfu.wu,pbalapra,michael.kruse,jkoo,bvideau,hovland,vtaylor\}@anl.gov}
\and
\linebreakand 
\IEEEauthorblockN{Brad Geltz, Siddhartha Jana}
\IEEEauthorblockA{Intel Corporation, Hillsboro, OR 97124  \\
            Email: \{brad.geltz,siddhartha.jana\}@intel.com} 
\and
\IEEEauthorblockN{Mary Hall}
\IEEEauthorblockA{University of Utah, Salt Lake City, UT 84103 \\ 
Email: mhall@cs.utah.edu}
}

%\fi

\maketitle

\thispagestyle{plain}
\pagestyle{plain}

\begin{abstract}
 As we enter the exascale computing era, efficiently utilizing power and optimizing the performance of scientific applications under power and energy constraints has become critical and challenging. We propose a low-overhead autotuning framework to autotune performance and energy for various hybrid MPI/OpenMP scientific applications at large scales and to explore the tradeoffs between application runtime and power/energy for energy efficient application execution, then use this framework to autotune four ECP proxy applications---XSBench, AMG, SWFFT, and SW4lite. Our approach uses Bayesian optimization with a Random Forest surrogate model to effectively search parameter spaces with up to 6 million different configurations on two large-scale production systems, Theta at Argonne National Laboratory and Summit at Oak Ridge National Laboratory. The experimental results show that our autotuning framework at large scales has low overhead and achieves good scalability.
Using the proposed autotuning framework to identify the best configurations, 
we achieve up to 91.59\% performance improvement, up to 21.2\% energy savings, and up to 37.84\% EDP improvement on up to 4,096 nodes.

\end{abstract}

%\begin{IEEEkeywords}
%autotuning, energy efficiency, performance optimization, hardware-software co-design
%\end{IEEEkeywords}

\input{introduction}

\input{background}

\input{systems-apps}

\input{framework}
\input{single}

\input{performance}
\input{energy}

\input{conclusions}

%\if 0
\section*{Acknowledgments}
This work was supported in part by DOE ECP PROTEAS-TUNE, in part by DOE ASCR RAPIDS2, and in part by NSF grant CCF-2119203.  We acknowledge the Argonne Leadership Computing Facility (ALCF) for use of Cray XC40 Theta under ALCF projects EE-ECP and Intel, and the Oak Ridge Leadership Computing Facility for use of Summit under the projects CSC383, MED106 and AST136. We also acknowledge Adrian Pope at ALCF for providing the SWFFT problem sizes. This material is based upon work supported by the U.S. Department of Energy, Office of Science, under contract number DE-AC02-06CH11357. Development of the GEOPM software package has been partially funded through contract B609815 with Argonne National Laboratory.
%\fi

%\IEEEtriggeratref{14}
%\newpage
\bibliographystyle{IEEEtran}
\bibliography{bibs}

%GAIL - don't  forget  the Govt license
\newpage

\end{document}

%% file: introduction.tex
\section{Introduction}

As we enter the exascale computing era, high performance, power, and energy management are key design points and constraints for any next generation of large-scale high-performance computing (HPC) systems~\cite{BB20, osti_powerstack, WM20}. Efficiently utilizing procured power and optimizing the performance of scientific applications under power and energy constraints are challenging for several reasons, including dynamic phase behavior, manufacturing variation, and increasing system-level heterogeneity. As the complexity of such HPC ecosystems (hardware stack, software stack, applications) continues to rise, achieving optimal performance and energy becomes a challenge. The number of tunable parameters that HPC users can configure at the system and application levels has increased significantly, resulting in a dramatically increased parameter space. Exhaustively evaluating all parameter combinations becomes very time-consuming. 
Therefore, autotuning for automatic exploration of the parameter space is desirable.

Autotuning is an approach that explores a search space of tunable parameter configurations of an application efficiently executed on an HPC system. Typically, one selects and evaluates a subset of the configurations on the target system and/or uses analytical models to identify the best implementation or configuration for performance or energy within a given computational budget. 
However, such methods are becoming too difficult in practice because of the hardware, software, and the application complexity. Recently, the use of advanced search methods that adopt mathematical optimization methods to explore the search space in an intelligent way has received significant attention in the autotuning community. Such a strategy, however, requires search methods to efficiently navigate the large parameter search space of possible configurations in order to avoid a large number of expensive application runs
%, while retaining just enough information 
to determine high-performance configurations or implementations. 
In this paper we propose a low-overhead machine learning (ML)-based autotuning framework to autotune four hybrid MPI/OpenMP Exascale Computing Project (ECP) proxy applications \cite{ECP}---XSBench \cite{XSB}, SWFFT \cite{SWF},  AMG \cite{AMG}, and SW4lite \cite{SW4L}---to improve their performance, energy, and energy delay product (EDP) on two large-scale HPC systems: Theta \cite{THETA} at Argonne National Laboratory (ANL), and Summit \cite{SUMMIT} at Oak Ridge National Laboratory (ORNL).
 
Traditional autotuning methods are built on heuristics that derive from automatically tuned BLAS libraries \cite{ATLAS}, experience \cite{TC02, CH06, GO10}, and model-based methods \cite{Chen05, TH11, BG13, FE17}. At the compiler level \cite{AK18}, ML-based methods are used for automatic tuning of the iterative compilation process \cite{OP17} and tuning of compiler-generated code \cite{TC09, MS14}.
Autotuning OpenMP codes has gone beyond loop schedules to look at parallel tasks and function inlining \cite{SJ19, KC14, MA11, Rose09}. Recent work on leveraging Bayesian optimization to explore the parameter space search shows the potential for autotuning on CPU systems \cite{WK20, WK21, LS21, RT21} and on GPU systems \cite{MS18, WN21}. Some recent work has used machine learning and sophisticated statistical learning methods to reduce the overhead of autotuning \cite{RB16, MA17, TN18, BQ19}. 
Most of these autotuning frameworks, however, are for autotuning on only a single or a few compute nodes using only performance as a metric. 

%For example,  GPTune \cite{LS21} has autotuned  MPI applications on up to 64 nodes with 2,048 cores with multitask learning using MPI, and Bayesian optimization was applied to increase the energy efficiency of a GPU cluster system \cite{MS18}. Our work is the first to autotune  hybrid MPI/OpenMP applications at large scales (up to 4,096 nodes with 262,144 cores), improving  performance and energy efficiency with low overhead and good scalability. \PB{commenting this because it was said in the related work}

This paper makes the following contributions.
 \begin{itemize}
 \item We propose a low-overhead autotuning framework ytopt to autotune various hybrid MPI/OpenMP applications at large scales.
 \item We use this ytopt framework to explore the tradeoffs between application runtime and power/energy for energy efficient application execution.
\item We use this framework to autotune four ECP proxy applications, namely XSBench, AMG, SWFFT, and SW4lite, using Bayesian optimization with a Random Forest surrogate model to effectively search parameter spaces with up to 6 million different configurations.
\item We demonstrate the effectiveness of our autotuning framework to tune the performance, energy, and EDP of these hybrid MPI/OpenMP applications on up to 4,096 nodes.
\item The experimental results show that our proposed autotuning framework at large scales has low overhead and good scalability, providing the best configuration for the best performance, energy saving, or EDP. Using the proposed autotuning framework to identify the best configurations, we achieve up to 91.59\% performance improvement, up to 21.2\% energy savings, and up to 37.84\% EDP improvement on up to 4,096 nodes.
\end{itemize}

The remainder of this paper is organized as follows. Section 2 discusses the background, challenges, and motivation of this study. Section 3 describes the systems and four ECP proxy applications used in this paper.  Section 4 proposes our autotuning frameworks for improving performance and energy at large scales. Section 5 discusses autotuning mixed pragmas on a single node. Section 6 presents autotuning performance at large scales. Section 7 illustrates autotuning energy and EDP at large scales. Section 8 summarizes this paper.

%% file: background.tex
\section{Background, Challenges, and Motivation}

%Considerable literature on autotuning exists. Balaprakash et al.\cite{BD18} surveyed the state of the practice in incorporating autotuned code into HPC applications; the authors highlighted insights from prior work and identified the challenges in advancing autotuning into wider and long-term use. Traditional autotuning methods are built on heuristics that derive from automatically tuned BLAS libraries \cite{ATLAS}, experience \cite{TC02, CH06, GO10}, and model-based methods \cite{Chen05, TH11, BG13, FE17}. At the compiler level \cite{AK18}, ML-based methods are used for automatic tuning of the iterative compilation process \cite{OP17} and tuning of compiler-generated code \cite{TC09, MS14}.
%Autotuning OpenMP codes has gone beyond loop schedules to look at parallel tasks and function inlining \cite{SJ19, KC14, MA11, Rose09}. Recent work on leveraging Bayesian optimization to explore the parameter space search shows the potential for autotuning on CPU systems \cite{WK20, WK21, LS21, RT21} and on GPU systems \cite{MS18, WN21}. Some recent work has used machine learning and sophisticated statistical learning methods to reduce the overhead of autotuning \cite{RB16, MA17, TN18, BQ19}.

Autotuning involves two critical requirements: (1) expression of a search space of implementations or configurations and (2) efficient navigation of the search space for identifying the optimal configuration. To address these two requirements, researchers have developed a number of autotuning frameworks  that interface with application codes, libraries, and compilers to generate code variants and measure their performance \cite{TC02,TH11,HN09,AK14, NR15,ZG15, RH17,CLTune,PG19,KernelTuner,HG20,YTO,WK20,WK21,KF20, KB21}. They presented the expression of a collection of parameters to be tuned and their corresponding possible values, and they generated possible configurations that may or may not be valid for evaluation. 

Two kinds of expressions of search space exist: vector space and tree space. 
Most autotuning frameworks present the search space in a vector space, that is, a fixed number of parameter knobs; these frameworks include OpenTuner \cite{AK14}, CLTune \cite{CLTune}, HalideTuner \cite{ZG15}, Orio \cite{HN09}, KernelTuner \cite{KernelTuner}, ATF \cite{RH17, RS21}, ytopt \cite{YTO, WK20, WK21}, GPTune \cite{LS21}, and Bliss \cite{RT21}. The successor of HalideTuner \cite{AM19} uses tree search to avoid the limitation of a vector search space but uses beam search to explore the space. ProTuner~\cite{HG20} further improves Halide schedule autotuning by replacing beam search with Monte Carlo tree search. The loop autotuner in Telamon also uses Monte Carlo tree search \cite{Telamon}. In the tradition of Halide, every level needs an assigned strategy, and a schedule where not all loops have an assigned strategy is considered incomplete. The viability of autotuning the search space for loop transformations was demonstrated; the approach involves the straightforward representation as either a tree or a directed acyclic graph using mctree \cite{MCTree, KF20, KB21}, and every loop is considered sequential until a pragma is added.

We classify  autotuning frameworks into four categories: (1) enumerate all possible parameter configurations, reject invalid ones, and evaluate the valid ones \cite{KF20}; (2) enumerate only valid configurations~\cite{RH17, RS21};  (3) sample from the set of possible configurations, and reject invalid ones~\cite{HN09, NR15, SJ19} during the search; and (4) sample only valid configurations, and search over them \cite{WK20, WK21}. The ytopt autotuning framework belongs to Category 4, which overcomes the ineffectiveness of Category 3 by generating valid samples and addresses the limitations of Categories 1 and 2, where enumerating all possible configurations can be computationally expensive for large number of parameters. However, the ytopt framework autotuned the applications only on a single computer node. Can we extend the ytopt framework to autotune MPI/OpenMP scientific applications at large scales so that we can identify the best configuration for running these applications on large-scale HPC systems efficiently? This is the main motivation of our work in this paper.
% MK: (4) cannot express unknown validity constraints (compile failure, runtime crash, ...)? Seems quite limiting 

Kruse and Finkel \cite{KF19} implemented a prototype of user-directed loop transformations using LLVM Clang \cite{clang} and Polly \cite{POLL} with additional loop transformation pragmas such as loop reversal, loop interchange, tiling, and array packing in the ECP SOLLVE project \cite{SOLL}. 
Multiple pragmas can be composed even to the same loop and every transformation addresses different and often contradicting concerns, such as maximizing parallelism, spatial and temporal memory locality, but minimizing bandwidth and overhead. Hence there is a need to determine how to efficiently combine them to optimize an application.

In  our recent work \cite{WK20, WK21} an autotuning framework ytopt was developed to leverage Bayesian optimization with four supervised machine learning methods---Random Forests, Gaussian Process Regression, Extra Trees, or Gradient-boosted Regression Trees---to explore the search space and identify more-promising regions\xingfu{, and we found the Random Forests performed the best}. This autotuning framework was used to identify the optimal combination of the Clang loop pragma parameters, with the aim of improving the performance of six PolyBench benchmarks \cite{YP16} and tuning the hyperparameters of a deep learning application MNIST on a single compute node. 

Most of autotuning frameworks mentioned above were for autotuning on a single or a few compute nodes. 
Recently, new autotuning frameworks are emerging for multi-node autotuning.  
For example, GPtune \cite{LS21} autotuned some MPI applications on up to 64 nodes with 2,048 cores with multitask learning using MPI, and Bayesian optimization was applied to increase the energy efficiency of a GPU cluster system \cite{MS18}. Current large-scale HPC systems such as Theta \cite{THETA} at ANL
%of approximately 12 petaflops peak performance at Argonne National Laboratory 
and Summit \cite{SUMMIT} 
%of approximately 200 petaflops peak performance 
at ORNL have  complex system architectures and software stacks with many tunable parameters that may affect the system performance and energy. How can we identify the best combination of these parameters for the best system performance or the lowest system energy consumption? Application developers and users often rely on these systems with the default configurations setup by the vendors to run their applications. How efficiently are these applications executed? 
%Answering these questions is difficult. 
Can we develop a low-overhead framework to autotune large-scale applications for performance and energy on large-scale HPC production systems such as Summit and Theta? 

The answer to these questions is "Yes, we can.'' Specifically, in this paper we demonstrate a new state of practice by applying autotuning approach to optimize performance and energy of hybrid MPI/OpenMP scientific applications on up to 4,096 nodes on these systems. 

%This is the focus of this paper. 

%% file: systems-apps.tex
\section{Systems and ECP Proxy Applications}
In this section we discuss the HPC system platforms and four ECP proxy applications \cite{ECP} used in our experiments.

We conduct our experiments on the Cray XC40 Theta \cite{THETA} of approximately 12 petaflops peak performance at Argonne National Laboratory and the IBM Power9 heterogeneous system Summit \cite{SUMMIT} of approximately 200 petaflops peak performance at Oak Ridge National Laboratory. In this section, we briefly describe their specifications shown in Table~\ref{tab:sys}. 

\textbf{Theta:}
Theta has 4,392 Cray XC40 nodes. Each node has 64 compute cores (one Intel Xeon Phi Knights Landing (KNL) 7230 with the thermal design power (TDP) of 215 W), shared L2 cache of 32 MB (1 MB L2 cache shared by two cores), 16 GB of high-bandwidth in-package memory Multi-Channel DRAM (MCDRAM), 192 GB of DDR4 RAM, and a 128 GB SSD. MCDRAM can be configured as a shared last level cache L3 (cache mode) or as a distinct NUMA node memory (flat mode) in or somewhere in between. The default memory mode is the cache mode. The Cray XC40 system uses the Cray Aries dragonfly network with user access to a Lustre parallel file system with 10 PB of capacity and 210 GB/s bandwidth. 

In this work, we use GEOPM \cite{ES17} to measure node energy consumption on Theta. The power sampling rate used is approximately 2 samples per second (default). We conduct all autotuning experiments in performance and energy with the cache mode. The compilers on Theta are CrayPE 2.6.5 (default) and clang 14 installed \cite{SOLL}. The aprun command is used to specify to ALPS (Application Level Placement Scheduler) the resources and placement parameters needed for the application at application launch on Theta.  

\textbf{Summit:}
Summit has 4,608 IBM Power System AC922 nodes. Each node contains two IBM POWER9 processors with 42 cores and six NVIDIA Volta V100 accelerators. Each node has 512 GB of DDR4 memory for use by the POWER9 processors and 96 GB of high-bandwidth memory (HBM2) for use by the accelerators. Additionally, each node has 1.6 TB of nonvolatile memory that can be used as a burst buffer. Summit is connected to an IBM Spectrum Scale filesystem providing 250 PB of storage capacity with a peak write speed of 2.5 TB/s. For each Summit node, the TDP of each Volta GPU is 300 W, and the TDP of each Power9 is 190 W. The power consumption of each Summit node is 2,200 W. Although we use the NVIDIA System Management Interface (nvidia-smi) \cite{NSMI} to measure power consumption for each GPU, the power measurement for IBM Power9 is not available to the public. Therefore, we autotune only performance of HPC applications on Summit. The compilers on Summit are gcc 9.1.0 (default) and nvhpc 21.3. The jsrun command is used for managing an allocation that is provided by an external resource manager within IBM Job Step Manager (JSM) software package on Summit. 

\begin{table}
\center
\caption{System Platform Specifications and Tools}
\begin{tabular}{c}
  \includegraphics[width=.80\linewidth]{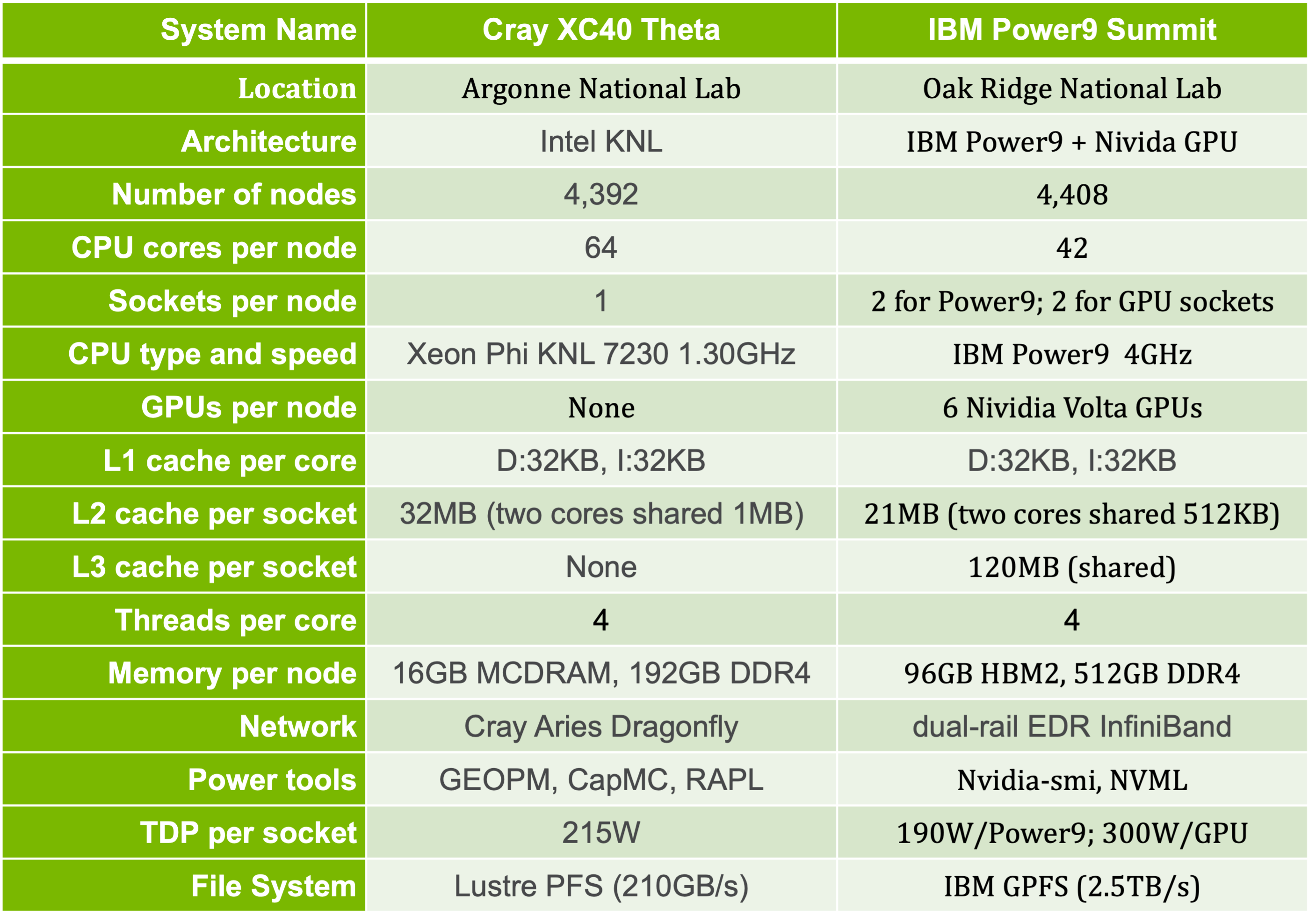}
  \end{tabular}
\label{tab:sys}       
\end{table}  

\subsection{ECP Proxy Applications}
%In high-performance computing, ECP proxy applications \cite{ECP} are small, simplified codes that allow application developers to share important features of large applications without forcing collaborators to assimilate large and complex code bases. 
In this section we discuss four hybrid MPI/OpenMP ECP proxy applications for our experiments: XSBench \cite{XSB}, SWFFT \cite{SWF}, AMG \cite{AMG}, and SW4lite \cite{SW4L}.

\subsubsection{Weak-Scaling Applications}
We discuss the three weak-scaling ECP proxy applications XSBench, SWFFT, and AMG.

XSBench \cite{XSB} is a mini-app representing a key computational kernel of the Monte Carlo neutron transport algorithm and represents the continuous energy macroscopic neutron cross section lookup kernel. It serves as a lightweight stand-in for full neutron transport applications like OpenMC \cite{OMC}. This code provides a much simpler and more transparent platform for determining performance benefits resulting from a given hardware feature or software optimization.  XSBench provides an MPI mode which runs the same code on all MPI ranks simultaneously with no decomposition across ranks of any kind, and all ranks accomplish the same work. It is an embarrassingly parallel implementation. It supports history-based transport (default):  parallelism is expressed over independent particle histories, with each particle being simulated in a serial fashion from birth to death; and event-based transport: parallelism is instead expressed over different collision (or "event") types.  XSBench is the hybrid MPI/OpenMP code written in C and supports OpenMP offload. The OpenMP offload implementation only supports the event-based transport. The problem size is large as default. 

SWFFT \cite{SWF} is to test the Hardware Accelerated Cosmology Code (HACC) 3D distributed memory discrete fast Fourier transform (FFT) with one forward FFT and one backward FFT. It assumes that global grid will originally be distributed between MPI ranks using a 3D Cartesian communicator. That data needs to be re-distributed to three 2D pencil distributions in turn in order to compute the double-precision FFTs along each dimension. SWFFT is the hybrid MPI/OpenMP code written in C++ and C and requires the cubic number of MPI ranks and FFTW3 (double precision, OpenMP version) installed. We configure it as weak scaling case. The problem size is 4096x4096x4096 for 4096 MPI ranks. We also set the number of run tests 2.

AMG \cite{AMG} a parallel algebraic multigrid solver for linear systems arising from problems on unstructured grids and builds linear systems for various 3-dimensional problems. Parallelism is achieved by data decomposition. AMG achieves this decomposition by simply subdividing the grid into logical X x Y x Z (in 3D) chunks of equal size. It is the hybrid MPI/OpenMP code written in C. The problem size is the 3D Laplace problem "-laplace -n 100 100 100 -P X Y Z".
This will generate a problem with 1,000,000 grid points per MPI process with a domain of the size 100*X x 100*Y x 100*Z. 

\subsubsection{Strong-Scaling Application}

SW4lite \cite{SW4L} is a bare bone version of SW4 \cite{SP12, SB18} (Seismic Waves, 4th order accuracy) intended for testing performance in a few important numerical kernels of SW4.  SW4 implements substantial capabilities for 3-D seismic modeling with a free surface condition on the top boundary, absorbing super-grid conditions on the far-field boundaries, and an arbitrary number of point force and/or point moment tensor source terms. It uses a fourth order in space and time finite-difference discretization of the elastic wave equations in displacement formulation. The large problem LOH.1-h50 is from the SCEC (Southern California Earthquake Center) test suite \cite{Day01}. It sets up a grid with a spacing h (=50) over a domain (X x Y x Z) 30000 x 30000 x 17000. It will run from time t=0 to t=9. The material properties are given by the block commands. They describe a layer on top of a half-space in the z-direction. A single moment point source is used with the time dependency being the Gaussian function. SW4lite is the hybrid MPI/OpenMP code written in C++ and Fortran90. In \cite{WU21}, performance and energy of SW4lite were optimized for the improved version. We use the improved version to define the parameter space for autotuning.

\subsubsection{Compiling Time for Each Application}

Table~\ref{tab:cp} shows the average compiling time (in seconds) for each ECP proxy application on Theta and Summit. We measured the compiling time for each application five times to get the average compiling time. We observe that the compiling time for SW4lite is 162.066 s on Theta and 58 s on Summit. This really impacts the autotuning wall-clock time in Step 4 shown in Figure \ref{fig:pf}. Because of loading the NVidia nvhpc module to compile the XSBench OpenMP offload version for using GPUs on Summit, it takes 4.645 s, which is much larger than that on Theta.

\begin{table}[ht]
\center
\caption{Compiling time (s) on Theta and Summit}
\begin{tabular}{|r|c|c|c|c|}
\hline
System & XSBench & SWFFT & AMG & SW4lite  \\
\hline
Theta &  2.021 & 3.494  & 2.825   &   162.066 \\
\hline
Summit & 4.645  & 3.781  & 2.757 &  58.000 \\
\hline
\end{tabular}
\label{tab:cp}
\end{table}

%% file: framework.tex
\section{Proposed Autotuning Frameworks in Performance or Energy at Large Scales}
In this section we extend the ytopt autotuning framework to autotune the hybrid MPI/OpenMP applications at large scales on the ANL Theta and ORNL Summit using the metrics such as performance, energy, and EDP, where the application runtime is the primary performance metric; energy consumption captures the tradeoff between the application runtime and power consumption; and EDP captures the tradeoff between the application runtime and energy consumption. 
%The aprun command is used to specify to the Cray ALPS (Application Level Placement Scheduler) the resources and placement parameters needed for the application at application launch on Theta \cite{THETA}. The jsrun command is used to manage an allocation that is provided by an external resource manager within the IBM Job Step Manager (JSM) software package on Summit \cite{SUMMIT}. At a high level, aprun or jsrun is similar to mpiexec or mpirun or srun.

\subsection{Framework for Autotuning Performance at Large Scales}

\begin{figure}[ht]
\center
 \includegraphics[width=.5\textwidth]{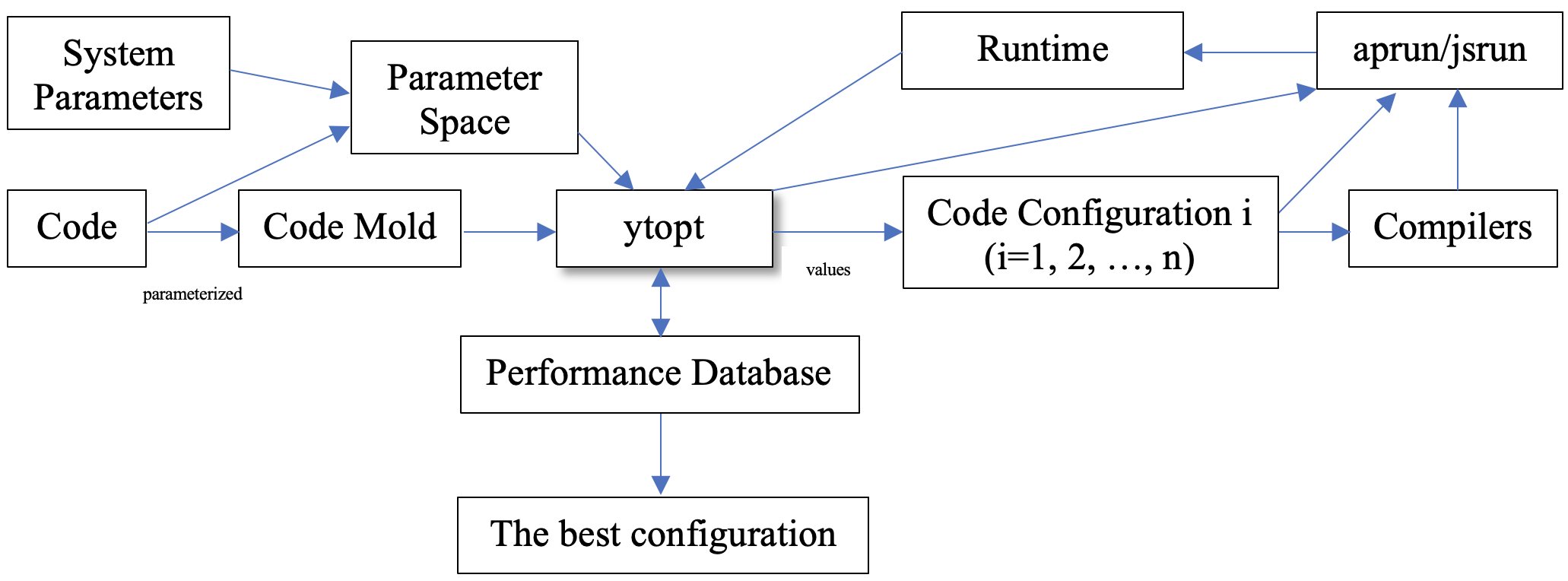}
 \setlength{\belowcaptionskip}{-8pt}
 \caption{Framework for Autotuning Performance}
\label{fig:pf}
\end{figure}

Figure \ref{fig:pf} presents the framework for autotuning various hybrid MPI/OpenMP applications in performance. The application runtime is the primary metric. We analyze an application code to identify the important tunable application and system parameters (OpenMP environment variables) to define the parameter space using ConfigSpace \cite{CFS} package.
We use the tunable parameters to parameterize an application code as a code mold. 
ytopt starts with the user-defined parameter space, the code mold, and user-defined interface that specifies how to evaluate the code mold with a particular parameter configuration. 

The search method within ytopt uses Bayesian optimization, where a dynamically updated Random Forest surrogate model that learns the relationship between the configurations and the performance metric, is used to balance exploration and exploitation of the search space. In the exploration phase, the search evaluates parameter configurations that improve the quality of the surrogate model, and in the exploitation phase, the search evaluates parameter configurations that are closer to the previously found high-performing parameter configurations. The balance is achieved through the use of the lower confidence bound (LCB) acquisition function that uses the surrogate models' predicted values of the unevaluated parameter configurations and the corresponding uncertainty values. \xingfu{The LCB acquisition function is defined in Equation \ref{eqn:lcb}. For the unevaluated parameter configuration $x_M^i$, the trained model $M$ is used to predict a point estimate (mean value) $\mu(x_M^i)$ and standard deviation $\sigma(x_M^i)$.
\begin{equation}
    a_{LCB}(x_M^i) = \mu(x_M^i) - \kappa\sigma(x_M^i)
    \label{eqn:lcb}
\end{equation}
where $\kappa \geq 0$ is a user-defined parameter that controls the tradeoff between exploration and exploitation. When $\kappa=0$ for pure exploitation, a configuration with the lowest mean value is selected. When $\kappa$ is set to a large value ($>1.96$) for pure exploration, a configuration with large predictive variance is selected. The default value $\kappa$ of is 1.96. Then the model $M$ is updated with this selected configuration.
%After the evaluation of the selected unseen configuration, it is used to update the model. Evaluation of such configurations results in improvement of the model $M$. 
}

\xingfu{The iterative phase of the proposed autotuning framework in performance has the following steps: 
\begin{itemize}
\item [Step1]  Bayesian optimization selects a parameter configuration for evaluation. 
\item [Step2] The code mold is configured with the selected configuration to generate a new code. 
\item [Step3]  Based on the value of the number of threads in the  configuration, the number of nodes reserved and the number of MPI ranks, aprun/jsrun command line for launching the application on the compute nodes is generated. 
\item [Step4] The new code is compiled with other codes needed to generate an executable. 
\item [Step5] The generated aprun/jsrun command line is executed to evaluate the application with the selected parameter configuration; the resulting application runtime is sent back to the search and recorded in the performance database. 
\end{itemize}

Steps 1--5 are repeated until the maximum number of code evaluations or the wall-clock time is exhausted for the autotuning run. }

In the rest of this paper,  the term \textbf{ytopt processing time} includes the time spent in the parameter space search, building the surrogate model, processing the selected configuration to generate a new code and the aprun/jsrun command line, compiling the new code, launching the application, and storing the configuration and performance in the performance database (except the application runtime). We use the term \textbf{ytopt overhead} to stand for the ytopt processing time minus the application compiling time. 

\xingfu{ytopt supports various application-tunable parameters, which impact the application performance but keep the program correctness.} The application-tunable parameters can be defined as variables, pragmas, pragma clauses, a statement, or a function (piece of code).  \xingfu{The combinations of these parameters with their ranges of values form a parameter space. This requires some knowledge about the applications and underlying systems.} For instance, \xingfu{we define "\#pragma omp parallel for" as a parameter before a loop to check how the application performance is affected with and without it.} 
%We  define the statement "MPI\_Barrier(MPI\_COMM\_WORLD);" as a parameter from a code still with its correctness to check how the application performance is affected with and without it. 
Table \ref{tab:sz} presents the parameter space for the four ECP proxy applications used in our experiments, where \textbf{system param.} stands for system parameters; \textbf{application param.} stands for unique application parameters because some of them are used repeatedly in the application code; and \textbf{space size} stands for the number of configurations for the parameter space. The system parameters in this paper mainly focus on OpenMP runtime environment variables \cite{OMP}:  OMP\_NUM\_THREADS, OMP\_PLACES, OMP\_PROC\_BIND, OMP\_SCHEDULE, and the additional OMP\_TARGET\_OFFLOAD. 

\xingfu{The selected application parameters which may impact performance for each application are described as follows.}
The two unique application parameters for XSBench are block size and additional "\#pragma omp parallel for." The five 
unique application parameters for XSBench-mixed (mixed Clang loop pragmas and OpenMP pragmas) are block size, Clang loop unrolling full, "\#pragma omp parallel for," and two tile sizes for 2D loop tiling. The  four unique application parameters for XSBench-offload are simd, device clause,  schedule for the OpenMP target pragmas, and "\#pragma omp parallel for." The one unique application parameter for SWFFT is "MPI\_Barrier(CartComm);". The  three unique application parameters for AMG are "\#pragma unroll(3)," "\#pragma unroll(6)," and "\#pragma omp parallel for." The four unique application parameters for SW4lite are ``\#pragma unroll (6)," "\#pragma omp parallel for," "\#pragma omp for nowait," and "MPI\_Barrier(MPI\_COMM\_WORLD);". Overall, we use the parameter spaces with up to 6,272,640 configurations for our experiments.  

\begin{table}[ht]
\center
\caption{Parameter Space for Each Application}
\begin{tabular}{|r|c|c|c|c|}
\hline
ECP Proxy Apps&  System param. &  Application param.  & Space size  \\
\hline
XSBench &  4 env. variables & 2  & 51,840    \\
\hline
XSBench-mixed &  4  env. variables &  5  & 6,272,640    \\
\hline
XSBench-offload &  5  env. variables & 4 & 181,440    \\
\hline
SWFFT & 4  env. variables  & 1  & 1,080  \\
\hline
AMG  & 4  env. variables  & 3  &   552,960 \\
\hline
SW4lite & 4 env. variables  & 4  &  2,211,840 \\
\hline
\end{tabular} 
\setlength{\belowcaptionskip}{-8pt}
\label{tab:sz}
\end{table}

\subsection{Framework for Autotuning Energy and EDP at Large Scales}
Efficiently utilizing the procured power and optimizing the  performance of scientific applications under power and energy constraints are important challenges in HPC. The HPC PowerStack \cite{4,osti_powerstack, BB20}---a global consortium of laboratories, vendors, and universities---has highlighted a design shift toward standardization of the HPC power-management software stack. This enables seamless integration of software solutions for managing the energy/power consumption of large-scale HPC systems. 
Based on the state of the art of the components available in the community for power and energy management \cite{GL16, GR19, 3, redfish, epajsrm, 5, GEOPM, RAPL}, a hierarchical strawman PowerStack design \cite{osti_powerstack} was proposed to manage power and energy at three levels of granularity:  system level,  job level, and  node level. This implies the need to put in place the following incrementally: 
(1) define policies that govern site-level requirements, a power-aware system Resource Manager (RM) / job scheduler, a power-aware job-level manager, and a power-aware node manager;
(2) define the interfaces between these layers to translate objectives at each layer into actionable items at the adjacent lower layer; 
and (3) drive end-to-end optimizations across different layers of the PowerStack.

In order to address these requirements, our recent  work  \cite{WM20} (a) surveyed the high-level objectives of the existing layer-specific tuning approaches at the different layers: system (i.e., cluster), job / application, and node, (b) defined the tunable parameters at each layer, and (c) proposed and discussed how to autotune the combination of different parameters at the distinct layers (parameter space) for an optimal solution (the smallest runtime or the lowest energy) under a system power cap. 

\begin{figure}[ht]
\center
 \includegraphics[width=.3\textwidth]{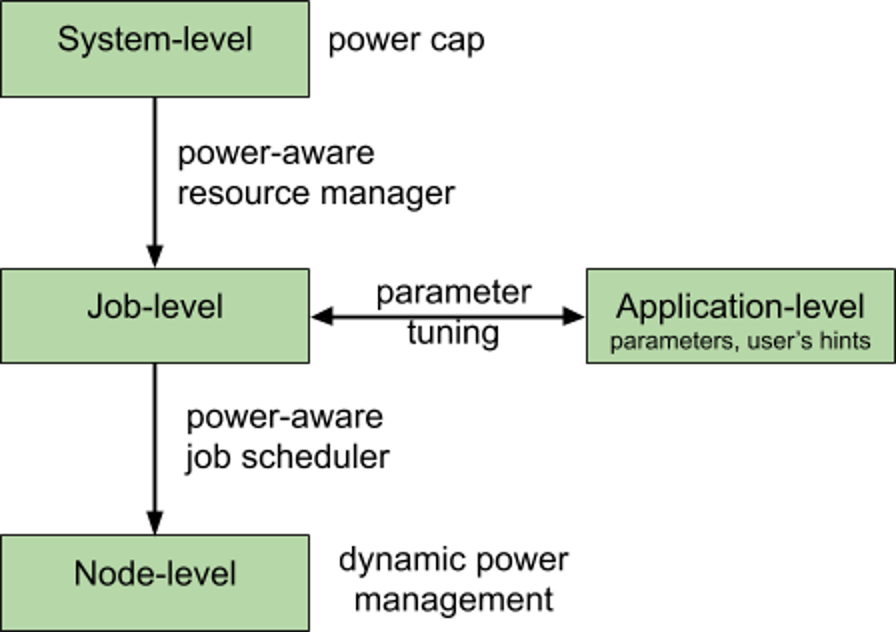}
 \setlength{\belowcaptionskip}{-8pt}
 \caption{PowerStack End-to-End Autotuning Framework}
\label{fig:ps}
\end{figure}

%\textbf{Research Question: } 
%How can we explore a holistic performance, power, and energy management software stack that can optimize the target 
%power- or energy-efficiency
%application-aware metric so that it can trade off power, energy, and time to solution in order to optimize the energy efficiency of an HPC system?
%GAIL - you said the next in Section II - althoug there you said you used the framework to identify optimal combination, not to optimize it
%Recent work \cite{WK21, WK20} developed an autotuning framework that leverages Bayesian optimization with four supervised machine learning methods to explore the parameter space search, and we used the autotuning framework to optimize the loop pragma parameters to improve the application performance. 
Based on our previous work on autotuning the performance, power, and energy of applications and systems \cite{WK21, WM21, WM20, WK20}, we propose a high-level end-to-end PowerStack autotuning framework for HPC systems shown in Figure \ref{fig:ps}. This diagram shows the interactions among four layers: system  level, job level, node level, and application level. 
To the best of our knowledge, however, a practical end-to-end autotuning component is still lacking that targets all four layers for the optimal solution. 
%Specifically, for DOE HPC platforms and simulations, each system node consists  not only of CPUs and GPUs but also of FPGAs and AI accelerators; and each simulation involves not only computation and communication but also surrogate model training, evaluation, and prediction. Thus, achieving an optimal system-level target metric while satisfying component-specific goals becomes challenging.

Our aim is to develop a practical framework to autotune all four layers of the PowerStack so that we can have a better understanding of the tunable parameters at each layer and interaction interfaces between layers and can identify potential new requirements in order to achieve energy efficiency. The process of autotuning in the layers (a) typically targets energy as the primary metric, (b) complies with the operating power constraint imposed on the layer, and (c) attempts to improve the management and orchestration of the available control parameters that affect the application and/or hardware performance. %The goal of this tuning is to enable an HPC system under power constraints to leverage feedback-driven interoperability between system resource managers, runtime systems, and applications to maximize system performance and energy efficiency. 
For the proposed framework, we integrate the existing job constraint-aware power/energy optimizer GEOPM (Global Extensible Open Power Manager)  \cite{ES17, GEOPM} at the job and node levels and the ytopt autotuning framework at the application level.

\begin{figure}[ht]
\center
 \includegraphics[width=.4\textwidth]{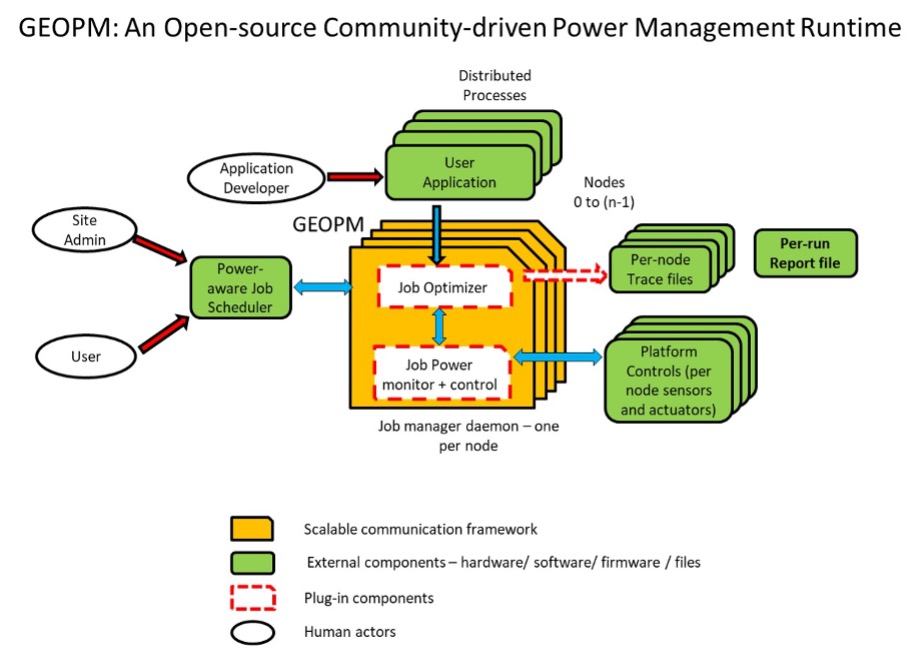}
 \setlength{\belowcaptionskip}{-8pt}
 \caption{High-Level Overview of GEOPM \cite{BB20}}
\label{fig:gpm}
\end{figure}

\begin{figure}[ht]
\center
 \includegraphics[width=.5\textwidth]{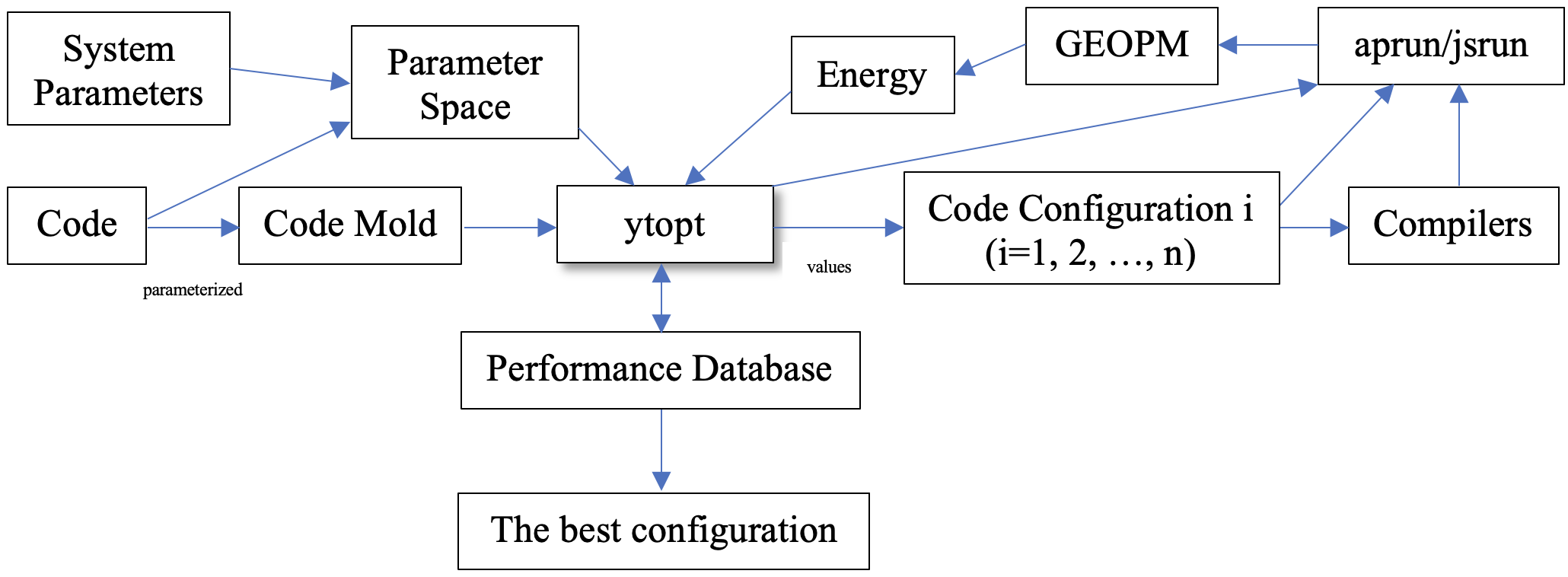}
 \setlength{\belowcaptionskip}{-12pt}
 \caption{Framework for Autotuning Energy}
\label{fig:en}
\end{figure}

GEOPM \cite{ES17, GEOPM} is a community-driven, cross-platform, open source, job-level power management framework.
It provides multiple interfaces to enable interoperability with external HPC software components such as enabling job schedulers and resource managers to drive job-aware system-wide power efficiency improvements in Figure \ref{fig:gpm}. %GEOPM is a part of the OpenHPC consortium and provides monitoring capabilities as well as control agents to optimize for time-to-solution by leveraging techniques from learning and control systems.
GEOPM enables control and monitoring of hardware/software knobs across multiple platforms and architectures such as leveraging multiple power and performance knobs like Intel's hardware power-limiting capability (RAPL \cite{RAPL_SDM}) for achieved CPU frequency and instructions retired.
Because the latest version of GEOPM (1.x) is installed on Theta but is not available on Summit \xingfu{because of special privilege requirement to access the low-level msr (model specific registers) counters} and the power measurement of Power9 is not available to the public on Summit, Figure \ref{fig:en} shows the proposed framework for autotuning energy and EDP of various hybrid MPI/OpenMP applications on Theta. The average node energy consumed by the application is the primary metric. 
%For the sake of simplicity, we do not consider the power-aware resource manager in this framework because of the use of the production system Theta.

This energy autotuning framework is similar to the performance framework with the five steps in Figure \ref{fig:pf}. Steps 1 and 2 are the same. There are some differences in Steps 3, 4, and 5. The GEOPM job launch script, geopmlaunch \cite{GEOPM}, queries and uses the OMP\_NUM\_THREADS environment variable to choose affinity masks for each process.  %The OMP\_NUM\_THREADS environment variable must be set before geopmlaunch is executed. 
The principal job of geopmlaunch to aprun is to set explicit per-process CPU affinity masks that will optimize performance while enabling the GEOPM controller thread to run on a core isolated from the cores used by the primary application. The geopmlaunch enables the GEOPM library to interpose on
MPI using the PMPI interface through the LD\_PRELOAD mechanism for unmodified binaries.  %In Step 3, ytopt sets the OMP\_NUM\_THREADS environment variable, and generates the aprun command line for application launch. In Step 4, the dynamic linking is required with the -dynamic flag for the compiling. In Step 5, ytopt uses the geopmlaunch to launch the aprun command line with the options "--geopm-ctl=pthread," which launches the controller as an extra pthread per node, and "--geopm-report=gm.report," which creates the summary report file gm.report about performance, power, and energy for each node to evaluate the application with the configuration. ytopt processes the summary report file from GEOPM to record the average node energy or EDP in the performance database. Steps 1--5 are repeated until the maximum number of code evaluations or the wall-clock time for the run on Theta.

The iterative phase of the proposed autotuning framework in energy has the following steps:
\begin{itemize}
\item [Steps] Steps 1 and 2 are the same as shown in Figure \ref{fig:pf}.
\item [Step3] ytopt sets the OMP\_NUM\_THREADS environment variable, and generates the aprun command line for application launch. 
\item [Step4] The dynamic linking is required with the -dynamic flag for the compiling. 
\item [Step5] ytopt uses the geopmlaunch to launch the aprun command line with the options "-{}-geopm-ctl=pthread," which launches the controller as an extra pthread per node, and "-{}-geopm-report=gm.report," which creates the summary report file gm.report about performance, power, and energy for each node to evaluate the application with the configuration. ytopt processes the summary report file from GEOPM to record the average node energy in the performance database. 
\end{itemize}

Steps 1--5 are repeated until the maximum number of code evaluations or the wall-clock time for the run on Theta.

%% file: single.tex
\section{Autotuning Mixed Pragmas on a Single Node}

In this section we apply the proposed framework in Figure \ref{fig:pf} to autotune XSBench with mixed pragmas on a single node of Theta and Summit. We use its OpenMP version on Theta and its OpenMP offload version on Summit.

\subsection{On Theta}

We modify the OpenMP version of XSBench by adding more OpenMP pragmas and Clang loop optimization pragmas, such as loop unrolling and tiling \cite{KF19}. % We use ytopt to autotune the mixed OpenMP loop pragmas and Clang loop pragmas version to achieve the best performance. Basically, 
We  integrate the new OpenMP pragmas with Clang loop pragmas as parameters to autotune the XSBench and to make sure that the result is verified. Note that we use the clang-14 compiler from SOLLVE LLVM \cite{SOLL} to compile the original and the mixed XSBench on Theta.

\if 0
{\scriptsize
\begin{verbatim}
cs = CS.ConfigurationSpace(seed=1234)
# number of threads
p0= CSH.OrdinalHyperparameter(name='p0', sequence=
['4','8','16','32','48','64','96','128','192','256'], default_value='64')
#block size for openmp dynamic schedule
p1= CSH.OrdinalHyperparameter(name='p1', sequence=
['10','20','40','50','64','80','100','128','160','200','256','400'], default_value='100')
#clang unrolling
p2= CSH.CategoricalHyperparameter(name='p2', 
choices=["#pragma clang loop unrolling full", " "], default_value=' ')
#omp parallel
p3= CSH.CategoricalHyperparameter(name='p3', 
choices=["#pragma omp parallel for", " "], default_value=' ')
# tile size for one dimension for 2D tiling
p4= CSH.OrdinalHyperparameter(name='p4', sequence=
['2','4','8','16','32','64','96','128','256','512','1024'], default_value='96')
# tile size for another dimension for 2D tiling
p5= CSH.OrdinalHyperparameter(name='p5', sequence=
['2','4','8','16','32','64','96','128','256','512','1024'], default_value='256')
# omp placement
p6= CSH.CategoricalHyperparameter(name='p6', 
choices=['cores','threads','sockets'], default_value='cores')
# OMP_PROC_BIND
p7= CSH.CategoricalHyperparameter(name='p7', 
choices=['close','spread','master'], default_value='close')
# OMP_SCHEDULE
p8= CSH.CategoricalHyperparameter(name='p8', 
choices=['dynamic','static','auto'], default_value='static')

cs.add_hyperparameters([p0, p1, p2, p3, p4, p5, p6, p7, p8])

\end{verbatim}
 }  
\fi
 
We use 9 parameters to define the following parameter space. The system runtime parameters are OMP\_NUM \_THREADS, OMP\_PLACES, OMP\_PROC\_BIND, and OMP\_SCHEDULE; the unique application parameters are block size for OpenMP dynamic schedule, Clang loop unrolling full, additional OpenMP parallel for, and two tile sizes for 2D loop tiling for a double nested loop (because this loop fails when parallelizing it in OpenMP).
 Because each Theta node has 64 cores with up to 4 threads per core, we choose 10 choices for OMP\_NUM\_THREADS in the range of 4 to 256 threads. The OpenMP specification includes many environment variables related to program execution \cite{OMP}. For the OMP\_PLACES, there are three options: cores (threads are allowed to float on cores), threads (threads are bound to specific logical processors), and sockets (threads are allowed to float on sockets). For the OMP\_PROC\_BIND, there are also three options:  close (threads placed consecutively), spread (threads spread equally on hardware), and master (threads placed on master  to enhance locality). OMP\_SCHEDULE allows specifying the schedule type (static, dynamic, or auto) with the default chunk size. 
For the block size (default 100 from the original code), we choose 12 choices in the range from 10 to 400. For the unrolling and additional OpenMP parallel for (4 in total), each has two choices with or without it. For two tile sizes for 2D loop tiling, we choose 11 choices in the range from 2 to 1,024 for each dimension. Therefore, the parameter space with total different configurations is 270*5808*4 =6,272,640, as shown in Table \ref{tab:sz}. 

\begin{figure}[ht]
    \centering
    \begin{subfigure}[t]{0.24\textwidth}
        \centering
        \includegraphics[width=\textwidth]{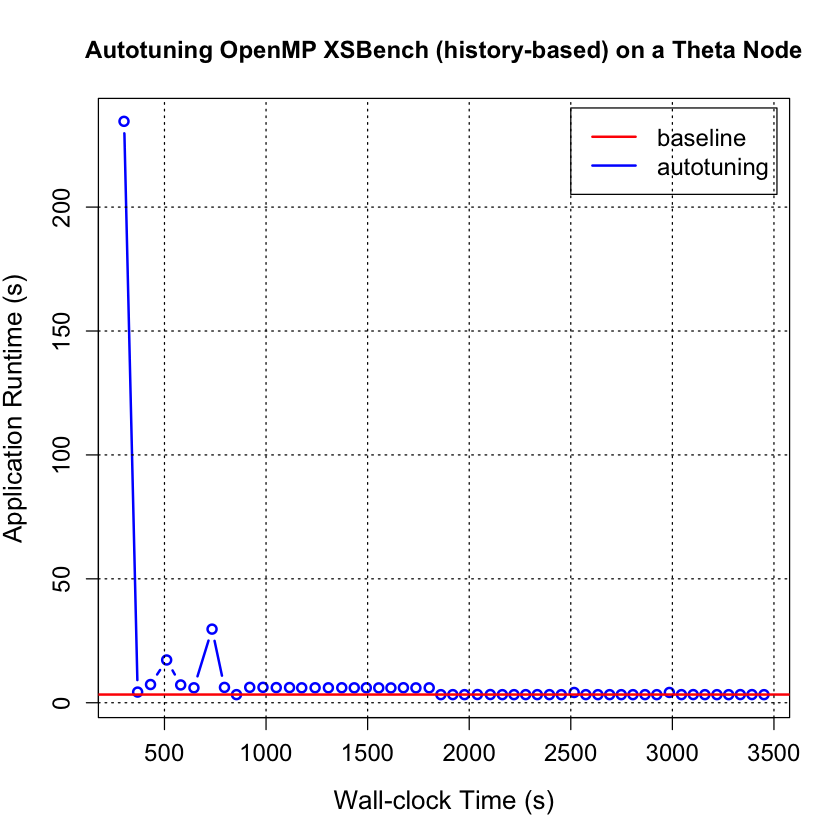}
        \subcaption{history-based}
        \label{fig:x1a}
    \end{subfigure}
    \begin{subfigure}[t]{0.24\textwidth}
        \centering
        \includegraphics[width=\textwidth]{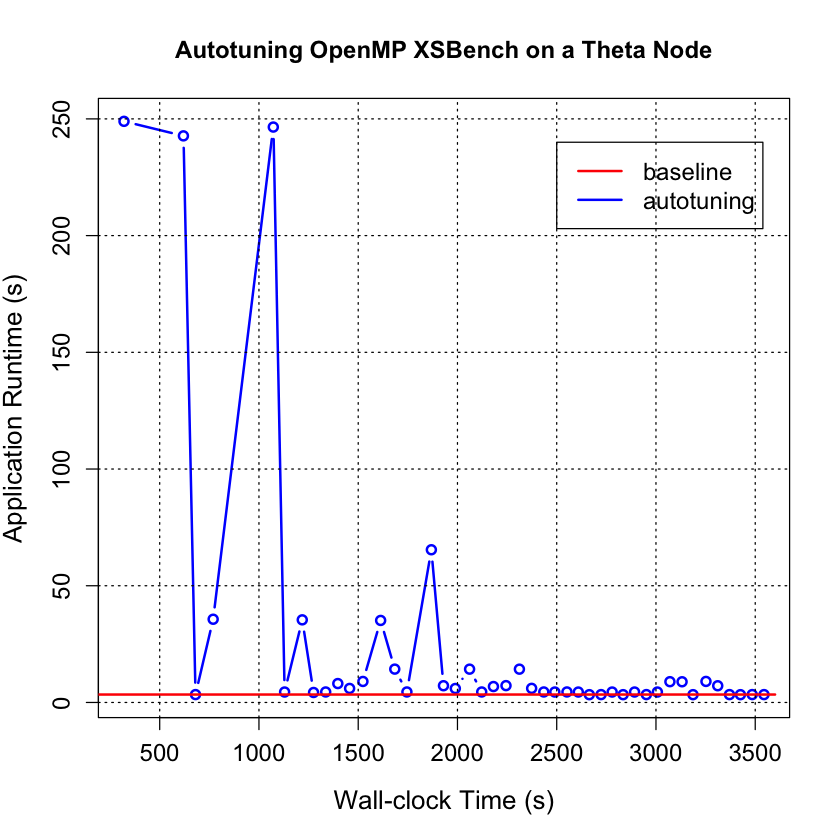}
        \subcaption{event based}
        \label{fig:x1e}
    \end{subfigure}
    \begin{subfigure}[t]{0.24\textwidth}
        \centering
        \includegraphics[width=\textwidth]{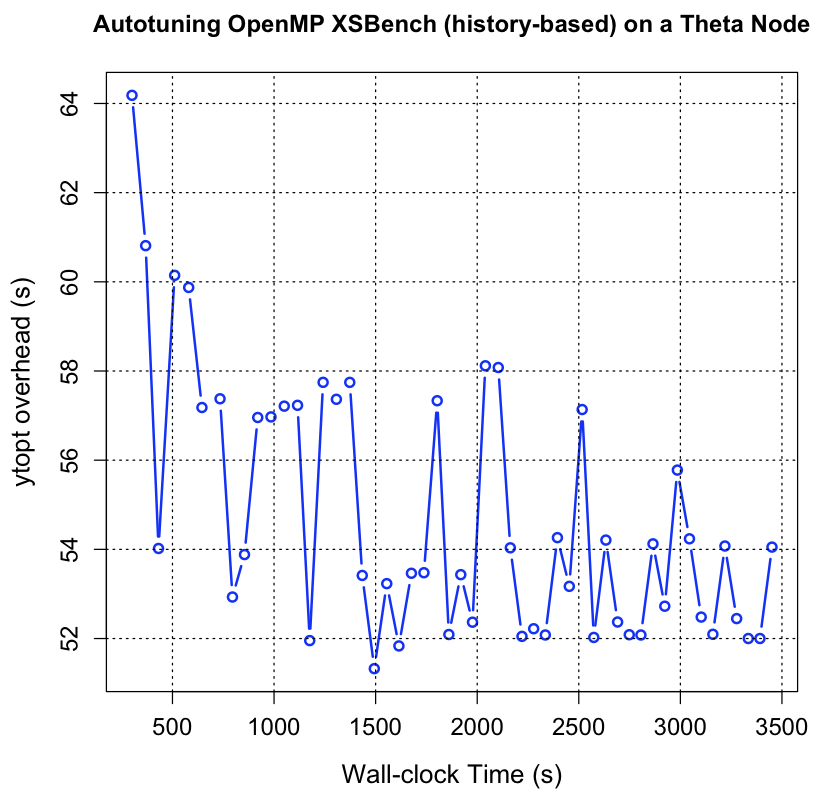}
        \subcaption{history based}
        \label{fig:x1o}
    \end{subfigure}
    \begin{subfigure}[t]{0.24\textwidth}
        \centering
        \includegraphics[width=\textwidth]{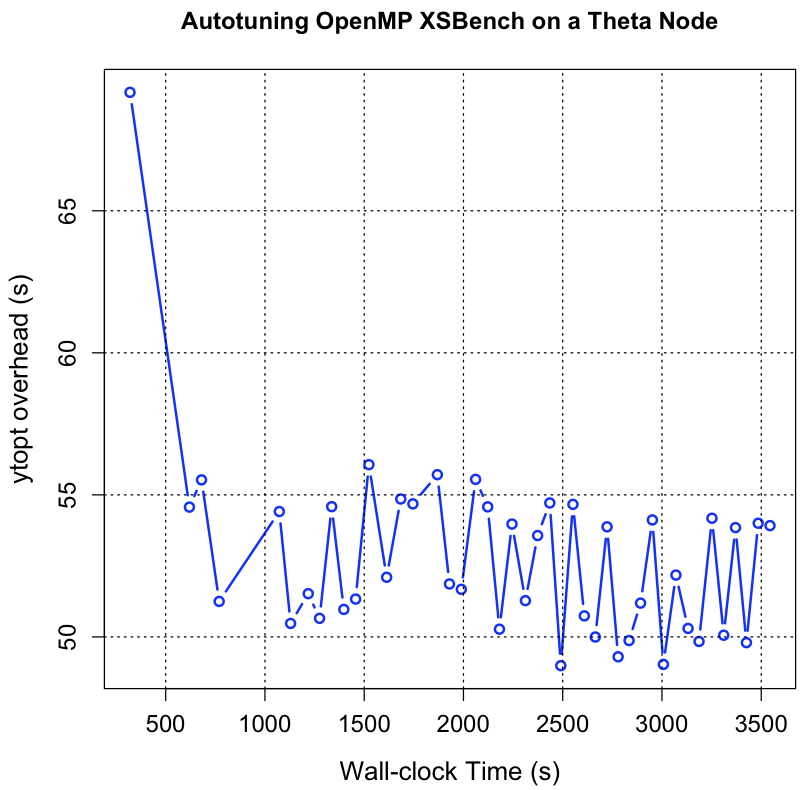}
        \subcaption{event based}
        \label{fig:x1eo}
    \end{subfigure}
    \setlength{\belowcaptionskip}{-8pt}
    \caption{Autotuning the Mixed-Pragmas Version of XSBench on a Theta Node}
    \label{fig:x1}
\end{figure}

%XSBench supports two simulations: history based and event based. We autotune both simulations. 
Figure~\ref{fig:x1a} shows the autotuning of the mixed pragmas version of XSBench (history based) on a single node, where wall-clock time stands for the time from the start of the autotuning to its end; the red line stands for the baseline runtime (3.31 s for the original code using 64 threads); and the blue line stands for the autotuning process over time. We achieve the best performance 3.262 s, and the search reaches the good region of the parameter space over time. Figure~\ref{fig:x1o} shows the ytopt overhead for each evaluation during the autotuning. 
The overhead is less than 65 s for the large parameter space.

Figure~\ref{fig:x1e} shows autotuning the mixed-pragmas version of XSBench (event based) on a single node. We observe the best performance 3.339 s (the baseline:  3.395 s for using 64 threads), and the search reaches the good region of the parameter space over time. Figure~\ref{fig:x1eo} shows the ytopt overhead for each evaluation over time. The overhead is between 49 s and 69.2 s for the large parameter space. We observe that the ytopt overhead for the first evaluation is the largest because it also includes setting the ytopt conda environment. Overall, the ytopt overhead is less than 70 s. 

%The best for event-based: 64,256,#pragma clang loop unrolling full,#pragma omp parallel for,32,1024,threads,spread,static,3.339
%The best for hisotry-based: 64,20,#pragma clang loop unrolling full,#pragma omp parallel for,256,64,cores,close,dynamic,3.262

\subsection{On Summit}

We use the OpenMP offload version of XSBench to autotune the application on a Summit node. %Specifically, we use it to add more OpenMP pragmas and optimization pragmas such as simd and schedule into the source code 
%GAIL - this "so that" seems a bit awkward   - do you mean "so that we can  use  ytopt? or do you  mean we  then use ytopt  to  autotune this  offload  version?
%so that we use ytopt to autotune it to achieve the best performance. 
We integrate the additional OpenMP pragmas with some clauses as parameters to autotune the XSBench and to make sure that the result is verified. The OpenMP offload version supports only the event-based simulation. In the rest of this paper, we use XSBench with the event-based simulation for our experiments. 

\if 0
{\scriptsize
\begin{verbatim}
cs = CS.ConfigurationSpace(seed=1234)
# number of threads (x4 for smt4)
p0= CSH.OrdinalHyperparameter(name='p0', 
sequence=['1','2','4','5','8','10','16','21','32','42''], default_value='42')
# omp placement
p1= CSH.CategoricalHyperparameter(name='p1', 
choices=['cores','threads','sockets'], default_value='cores')
# OMP_PROC_BIND
p2= CSH.CategoricalHyperparameter(name='p2', 
choices=['close','spread','master'], default_value='close')
#omp parallel
p3= CSH.CategoricalHyperparameter(name='p3', 
choices=["#pragma omp parallel for", " "], default_value=' ')
#omp parallel simd
p4= CSH.CategoricalHyperparameter(name='p4', 
choices=["simd", " "], default_value=' ')
#omp target device selected
p5= CSH.CategoricalHyperparameter(name='p5', 
choices=["device(offloaded_to_device)", " "], default_value=' ')
# OMP_SCHEDULE
p6= CSH.CategoricalHyperparameter(name='p6', 
choices=['dynamic','static','auto'], default_value='static')
#omp target schedule
p7= CSH.CategoricalHyperparameter(name='p7', choices=
["schedule(static,1)","schedule(static,2)","schedule(static,4)","schedule(static,8)",
"schedule(static,16)","schedule(static,32)", " "], default_value=' ')
#OMP_TARGET_OFFLOAD
p8= CSH.CategoricalHyperparameter(name='p8', 
choices=['DEFAULT','DISABLED','MANDATORY'], default_value='DEFAULT')

cs.add_hyperparameters([p0, p1, p2, p3, p4, p5, p6, p7, p8])
\end{verbatim}
 }  
 \fi
 
We use 9 parameters to define the following parameter space. The system runtime parameters are OMP\_NUM\_THREADS, OMP\_PLACES, OMP\_PROC\_ BIND, OMP\_SCHEDULE, and OMP\_TARGET\_OFFLOAD; the unique application parameters are additional ``\#pragma omp parallel for'', simd, device, and schedule clauses. Because each Summit node has 42 cores with up to 4 threads per core, we choose 10 choices for OMP\_NUM\_THREADS in the range from 4 to 168 threads. OMP\_TARGET\_OFFLOAD affects the behavior of execution on host and device including host fallback and provides three options: DEFAULT (try to execute on a GPU; if a supported GPU is not available, fall back to the host), DISABLED (do not execute on the GPU even if one is available; execute on the host), and MANDATORY (execute on a GPU or terminate the program). The simd clause is to create a team of threads to execute the loop in parallel using SIMD instructions. The device clause is to evaluate an assigned non-negative integer value less than the value of omp\_get\_num\_devices() (6 devices on a Summit node). schedule(static,1) for the OpenMP target pragmas is for memory access coalescing; scheduling a chunk size of 1 for each thread allows consecutive threads to access consecutive global memory locations. We choose six chunk sizes in the range from 1 to 32, adding one of them or without adding one as the total 7 choices for the parameter schedule. Therefore, the parameter space with all the different configurations is 810*56*4=181,440, as shown in Table~\ref{tab:sz}. 

\begin{figure}[ht]
    \centering
    \begin{subfigure}[t]{0.24\textwidth}
        \centering
        \includegraphics[width=\textwidth]{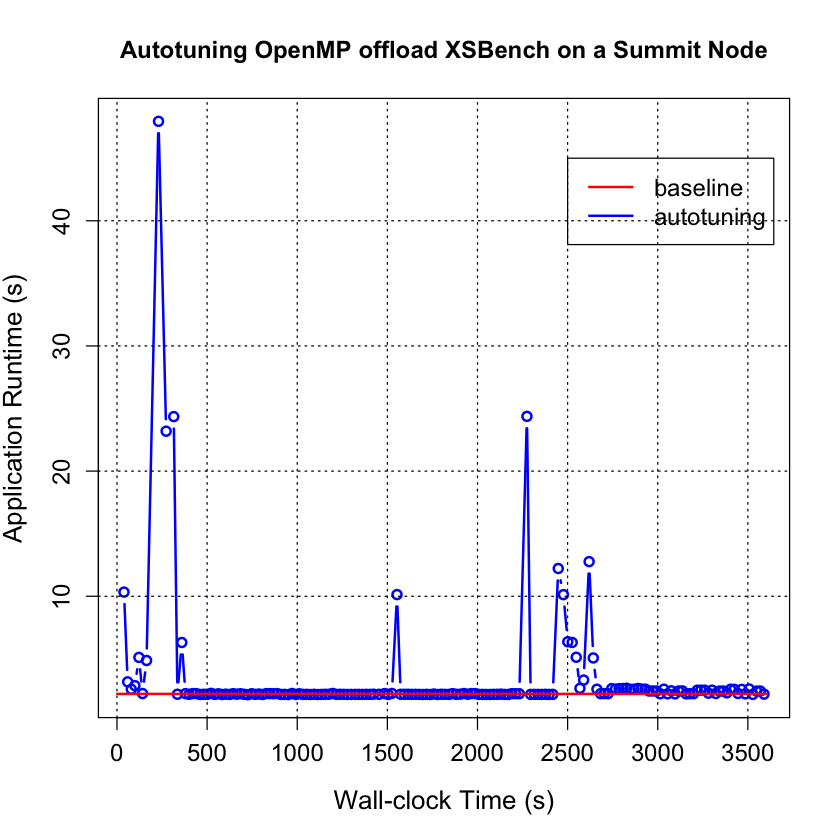}
        \subcaption{Application runtime}
        \label{fig:s1a}
    \end{subfigure}
    \begin{subfigure}[t]{0.24\textwidth}
        \centering
        \includegraphics[width=\textwidth]{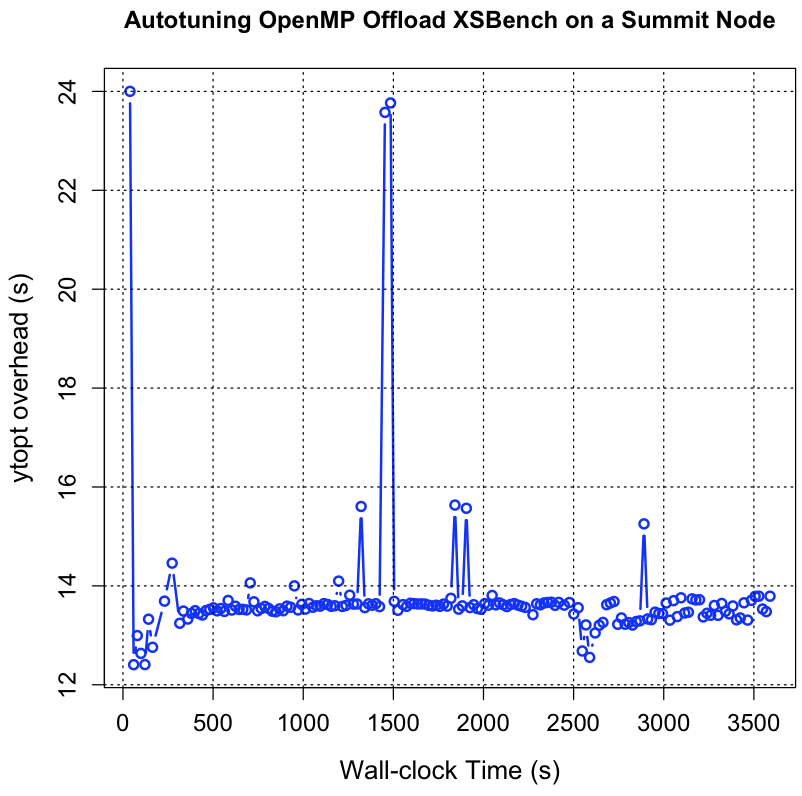}
        \subcaption{ytopt overhead}
        \label{fig:s1o}
    \end{subfigure}
    \setlength{\belowcaptionskip}{-8pt}
    \caption{Autotuning OpenMP Offload Version of XSBench (Event Based) on a Summit Node}
    \label{fig:x1s}
\end{figure}

Figure~\ref{fig:x1s} shows the autotuning of the OpenMP offload version of XSBench (event  based) on a single Summit node using 6 GPUs. We observe the best performance 2.138 s (the baseline: 2.20 s for using 6 GPUs and 168 threads), and the search reaches the good region of the parameter space over time in Figure~\ref{fig:s1a}. Figure~\ref{fig:s1o} shows the ytopt overhead for each evaluation during the autotuning. It is less than 24 s. This is much faster than on Theta.

%% file: performance.tex
\section{Autotuning Performance at Large Scales}

In this section we apply the proposed framework in Figure~\ref{fig:pf} to autotune the performance of four ECP proxy applications---XSBench, AMG, SWFFT, and SW4lit---on both Theta and Summit. To launch an application to compute nodes, Theta uses aprun, and Summit uses jsrun. The processor core on both Theta and Summit supports the simultaneous multithreading (SMT) level of 4 as default so that the number of threads per node is supported up to 256 on Theta and up to 168 on Summit. In Step 3 shown in Figure~\ref{fig:pf}, based on the value of the number of threads from the selected configuration, the number of nodes reserved, and the number of MPI ranks, ytopt generates the aprun/jsrun command line for application launch on compute nodes. For instance, we reserve 4,096 nodes with one MPI rank per node to run an application on Theta and Summit. 

On Theta we use the following algorithm to generate an aprun command line.
{\scriptsize
\begin{verbatim}
OMP_NUM_THREADS=n
if (n <= 64) {
 aprun -n 4096 -N 1 -cc depth -d n -j 1 application 
} else { if (n <= 128) {
  aprun -n 4096 -N 1 -cc depth -d n/2 -j 2 application 
 } else { if (n <= 192) {
     aprun -n 4096 -N 1 -cc depth -d n/3 -j 3 application
    } else {
       	aprun -n 4096 -N 1 -cc depth -d n/4 -j 4 application 
    }
  }
}
\end{verbatim}
}  

When we choose the number of threads n for each case, we make sure that n/2, n/3, or n/4 is integer on Theta. Then we use the algorithm to set the proper number of threads per core to generate the aprun command line.

On Summit we use the following algorithm to generate the jsrun command line. When the application uses 6 GPUs per node for the hybrid MPI/OpenMP offload application XSBench, we use the algorithm.
{\scriptsize
\begin{verbatim}
OMP_NUM_THREADS=n
jsrun -n4096 -a6 -g6 -c42 -bpacked:n/4 -dpacked  application 
\end{verbatim}
}  
When we choose the number of threads n, we make sure that n/4 is an integer because of the SMT level of 4 as default on Summit. We set one MPI rank per GPU and 42 cores per node for threads.

When the application uses only CPUs per node without any GPU for the hybrid MPI/OpenMP applications AMG, SWFFT, and SW4lite, we use the following algorithm to set one MPI rank per node and 42 cores per node for threads.
{\scriptsize
\begin{verbatim}
OMP_NUM_THREADS=n
jsrun -n4096 -a1 -g0 -c42 -bpacked:n/4 -dpacked  application 
\end{verbatim}
}  

For measuring the baseline performance for each application with a given problem size, we set the number of threads to 64 (which results in the best performance) on Theta and 168 threads (which also results in the best performance) on Summit to run the application under the default system configuration five times. Then we use the smallest application runtime as the baseline for the application. Notice that because of the limited node-hour allocations on Theta and Summit for our projects, we had to set most of the wall-clock times for autotuning runs at half an hour (1800 s). This limits the number of evaluations for different configurations during the autotuning. 
%GAI  - do you really need the next sentence>
%An evaluation of an application means that the application with a configuration is executed and its performance is evaluated.

\begin{figure}[ht]
    \centering
    \begin{subfigure}[t]{0.24\textwidth}
        \centering
        \includegraphics[width=\textwidth]{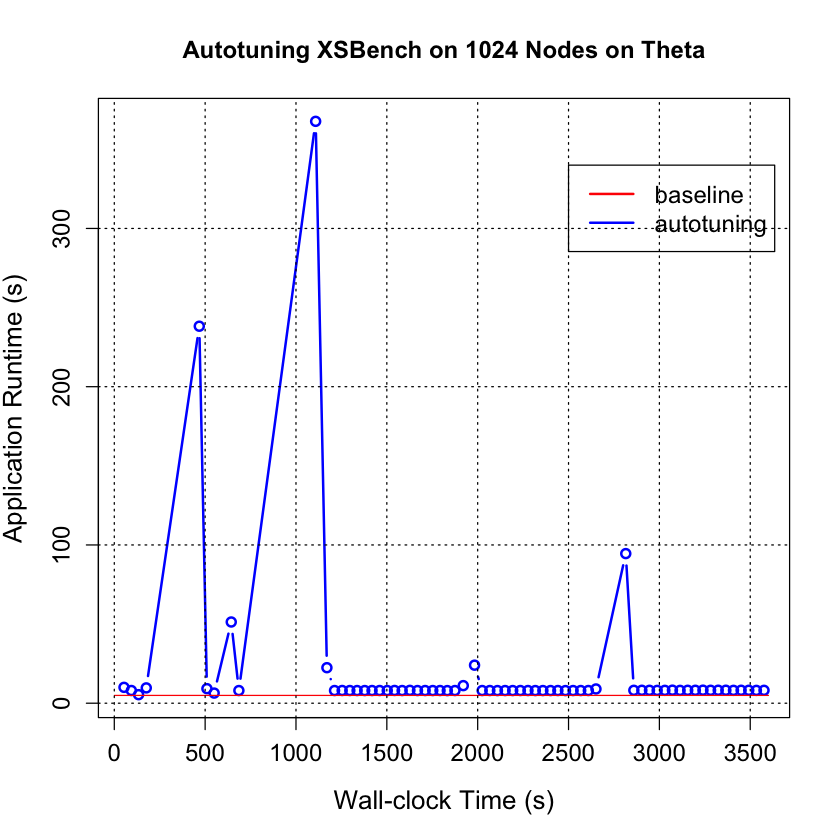}
        \subcaption{on 1024 nodes}
        \label{fig:x2a}
    \end{subfigure}
    \begin{subfigure}[t]{0.24\textwidth}
        \centering
        \includegraphics[width=\textwidth]{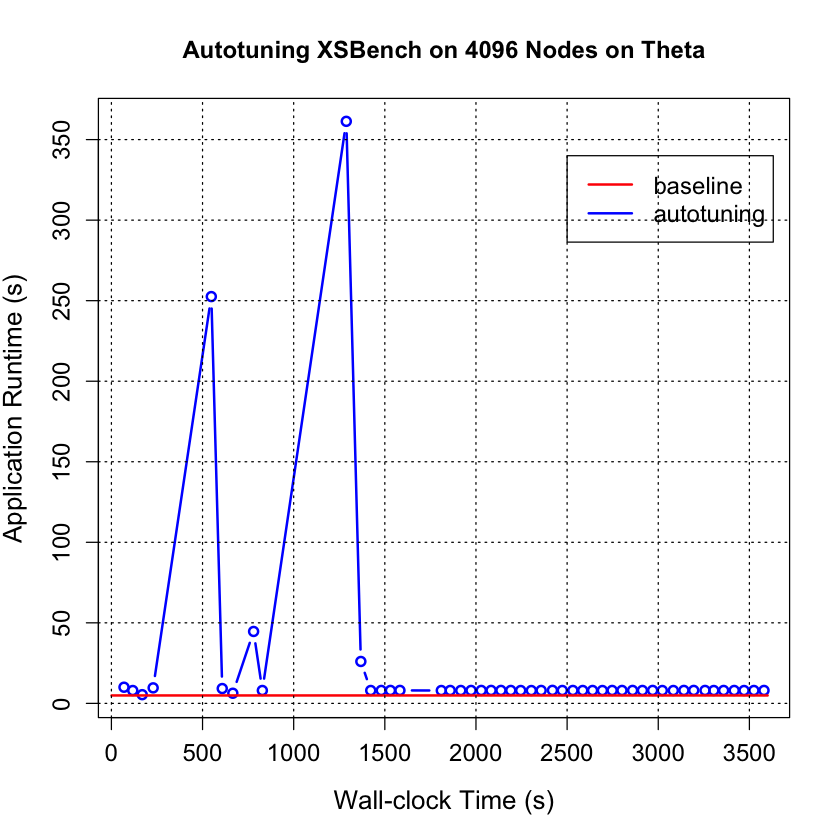}
        \subcaption{on 4096 nodes}
        \label{fig:x2b}
    \end{subfigure}
    \setlength{\belowcaptionskip}{-8pt}
    \caption{Autotuning XSBench at Large Scales on Theta}
    \label{fig:x2}
\end{figure}

Figure~\ref{fig:x2} shows autotuning MPI/OpenMP XSBench with the large problem size on 1,024 and 4,096 nodes on Theta. Because XSBench is weak scaling, both the autotuning processes are similar because of its embarrassingly parallel implementation of XSBench. We observe that the ytopt search reaches the good region of the parameter space over time that is close to that of the baseline. The ytopt overhead is similar to that in Figure~\ref{fig:x1eo} on Theta.

\begin{figure}[ht]
    \centering
    \begin{subfigure}[t]{0.24\textwidth}
        \centering
        \includegraphics[width=\textwidth]{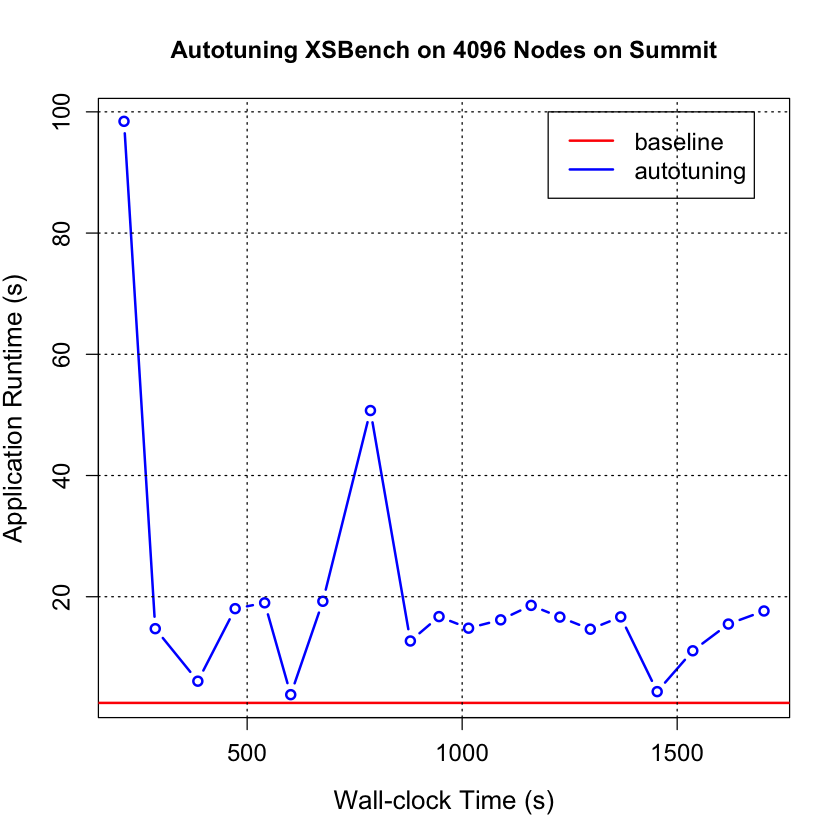}
        \subcaption{on 4096 nodes}
        \label{fig:x3a}
    \end{subfigure}
    \begin{subfigure}[t]{0.24\textwidth}
        \centering
        \includegraphics[width=\textwidth]{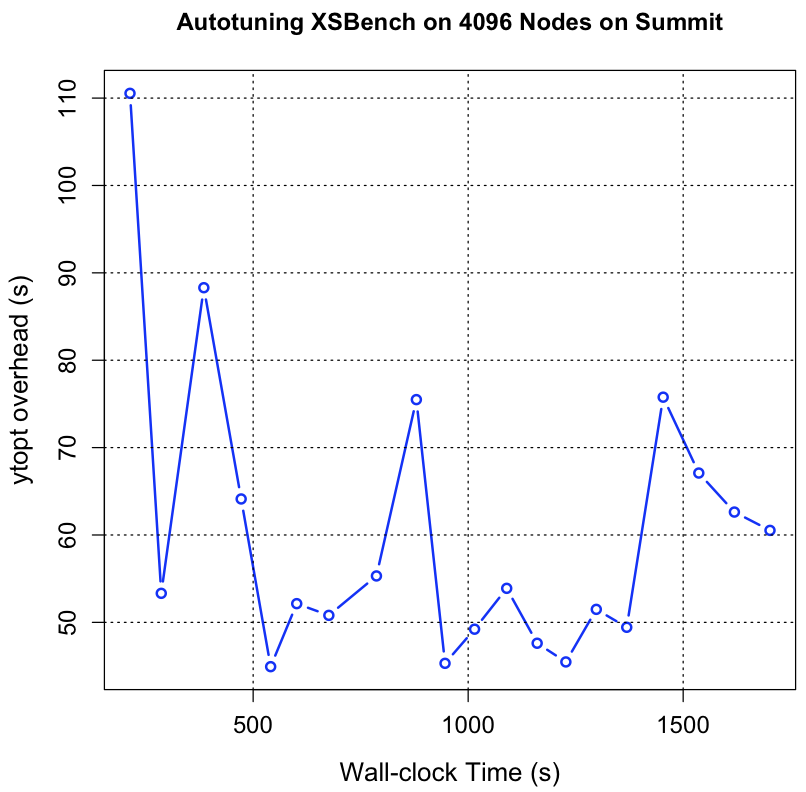}
        \subcaption{ytopt overhead}
        \label{fig:x3b}
    \end{subfigure}
    \setlength{\belowcaptionskip}{-8pt}
    \caption{Autotuning XSBench at Large Scale on Summit}
    \label{fig:x3}
\end{figure}

Figure~\ref{fig:x3} shows autotuning MPI/OpenMP offload XSBench using 6 GPUs per node and 1 MPI rank per GPU on 4096 nodes on Summit. We observe that the ytopt search gradually reaches the good region of the parameter space over time (baseline: . Because of the limited number of evaluations (20), however, it does not reach the optimal performance yet in Figure~\ref{fig:x3a}. Figure~\ref{fig:x3b} shows the ytopt overhead during the entire autotuning. Notice that the first ytopt overhead (111 s) also includes the time spent in setting the ytopt conda environment and loading the nvhpc module, however, most of the times are around 60 s.  The ytopt overhead is less than 111 s.

\begin{figure}[ht]
    \centering
    \begin{subfigure}[t]{0.24\textwidth}
        \centering
        \includegraphics[width=\textwidth]{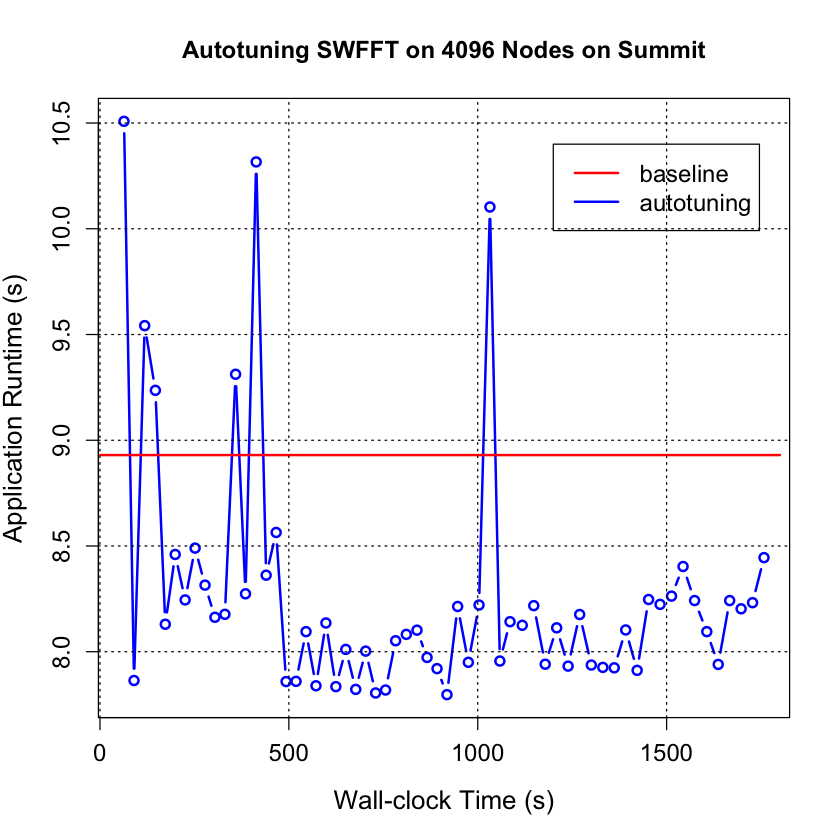}
        \subcaption{on 4096 nodes}
        \label{fig:s2a}
    \end{subfigure}
    \begin{subfigure}[t]{0.24\textwidth}
        \centering
        \includegraphics[width=\textwidth]{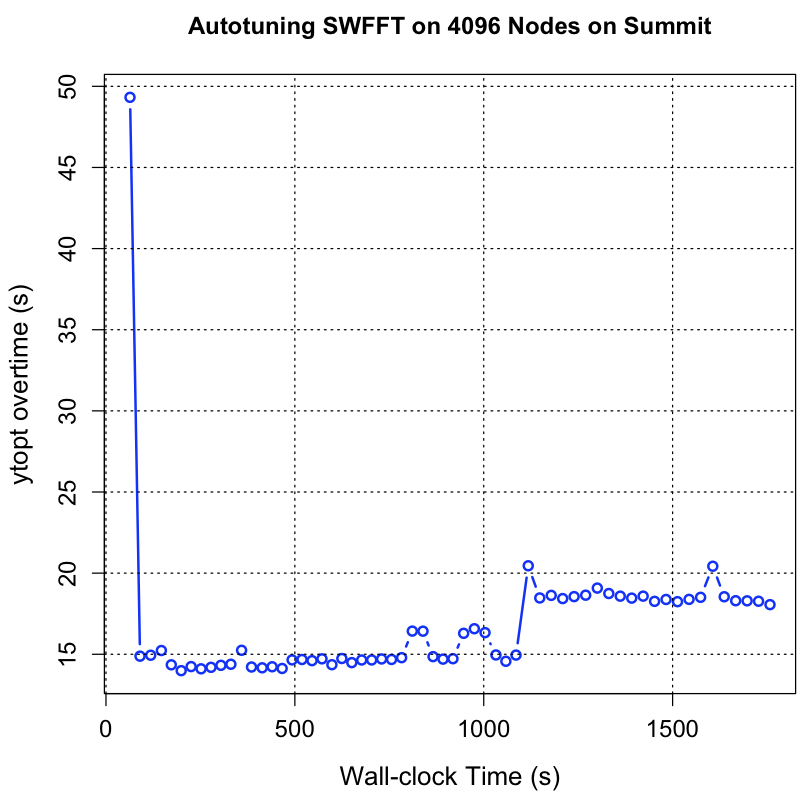}
        \subcaption{ytopt overhead}
        \label{fig:s2b}
    \end{subfigure}
    \setlength{\belowcaptionskip}{-8pt}
    \caption{Autotuning SWFFT at Large Scale on Summit}
    \label{fig:s2}
\end{figure}

SWFFT is weak scaling. Figure~\ref{fig:s2} shows autotuning SWFFT with a problem size of 3D grid 4096x4096x4096 on 4,096 nodes on Summit. We observe that the ytopt search reaches the good region of the parameter space over time with the smallest runtime of 7.797 s that is better than the baseline (8.93s) in Figure~\ref{fig:s2a}. This is 12.69\% performance improvement. The ytopt overhead is shown in Figure~\ref{fig:s2b}, and most of the times are around 20 s because of the small parameter space for SWFFT. So the ytopt overhead is less than 50 s.

\begin{figure}[ht]
    \centering
    \begin{subfigure}[t]{0.24\textwidth}
        \centering
        \includegraphics[width=\textwidth]{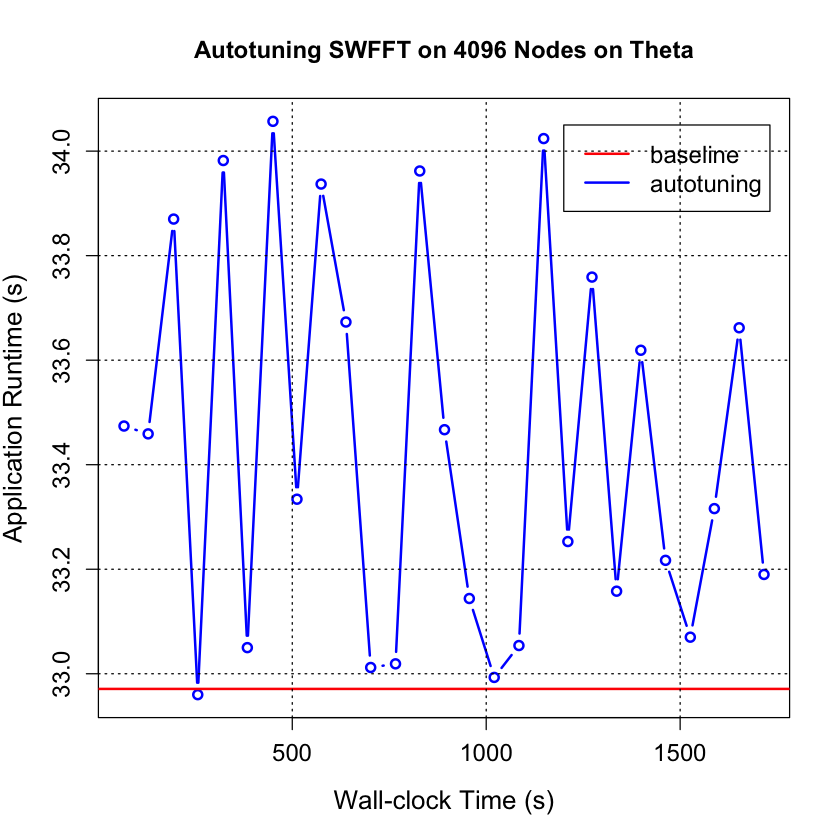}
        \subcaption{on 4096 nodes}
        \label{fig:s11a}
    \end{subfigure}
    \begin{subfigure}[t]{0.24\textwidth}
        \centering
        \includegraphics[width=\textwidth]{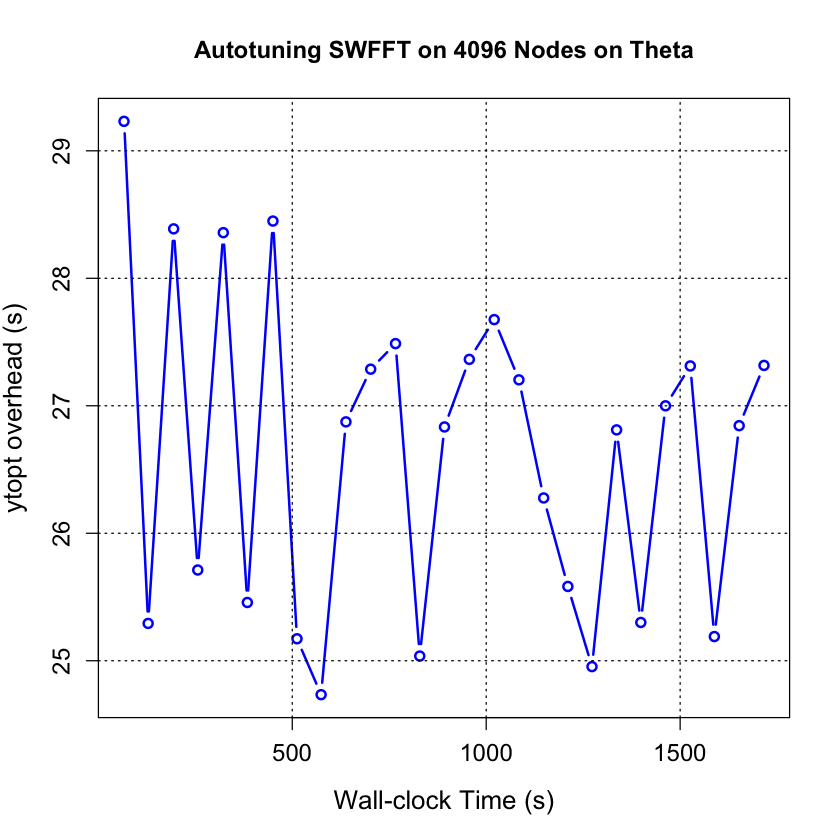}
        \subcaption{ytopt overhead}
        \label{fig:s11b}
    \end{subfigure}
    \setlength{\belowcaptionskip}{-8pt}
    \caption{Autotuning SWFFT at Large Scale on Theta}
    \label{fig:s11}
\end{figure}

Figure \ref{fig:s11} shows autotuning SWFFT with the same problem size on 4,096 nodes on Theta. We observe that the ytopt search reaches the good region of the parameter space over time that is close to the time of the baseline in Figure~\ref{fig:s11a}. The ytopt overhead is less than 30 s in Figure~\ref{fig:s11b}. 

\begin{figure}[ht]
    \centering
    \begin{subfigure}[t]{0.24\textwidth}
        \centering
        \includegraphics[width=\textwidth]{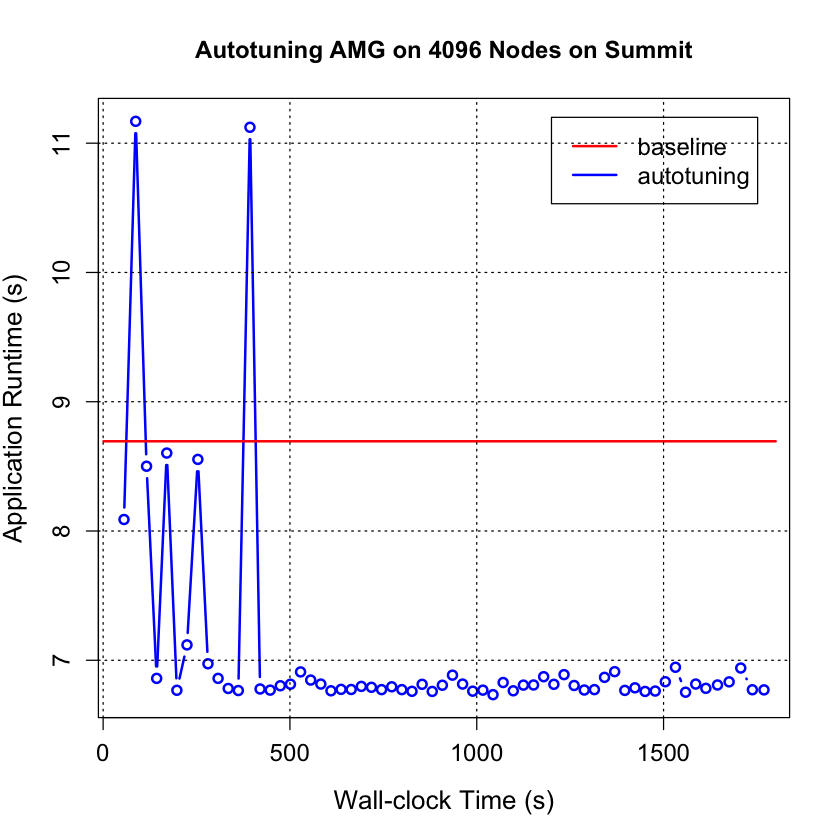}
        \subcaption{on 4096 nodes}
        \label{fig:a2a}
    \end{subfigure}
    \begin{subfigure}[t]{0.24\textwidth}
        \centering
        \includegraphics[width=\textwidth]{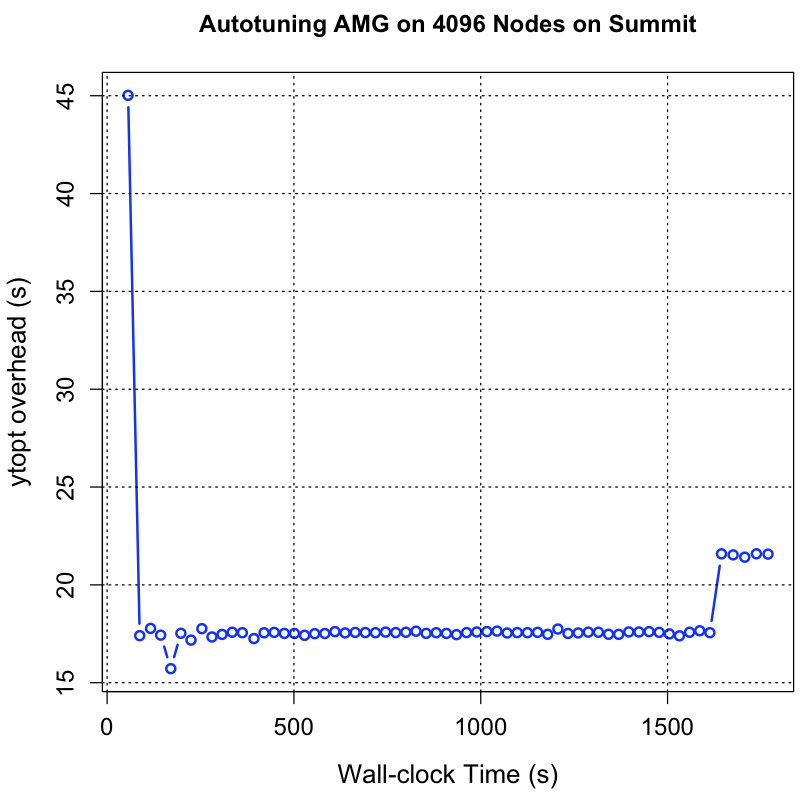}
        \subcaption{ytopt overhead}
        \label{fig:a2b}
    \end{subfigure}
    \setlength{\belowcaptionskip}{-8pt}
    \caption{Autotuning AMG at Large Scale on Summit}
    \label{fig:a2}
\end{figure}

AMG is weak scaling. Figure~\ref{fig:a2} shows autotuning of AMG on 4,096 nodes on Summit. We use the 3D Laplace problem "-laplace -n 100 100 100 -P 16 16 16" as the input, which means generating a problem with 1,000,000 grid points per MPI rank with a domain size 1600 x 1600 x 1600 on 4,096 nodes with 1 MPI rank per node and various numbers of threads per MPI rank. We observe that the ytopt autotuning reaches the best configuration with the smallest runtime of 6.734 s, which is much better than the baseline performance of 8.694 s in Figure~\ref{fig:a2a}. This is a 22.54\% performance improvement. Figure~\ref{fig:a2b} shows the ytopt overhead is less than 45 s. 
 
\begin{figure}[ht]
    \centering
    \begin{subfigure}[t]{0.24\textwidth}
        \centering
        \includegraphics[width=\textwidth]{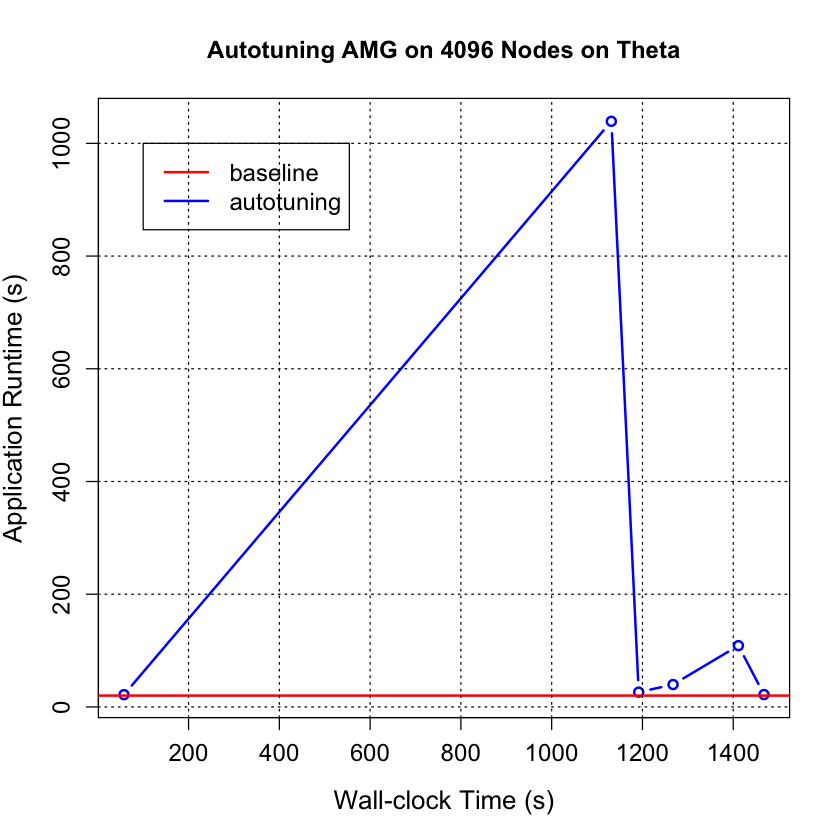}
        \subcaption{on 4096 nodes}
        \label{fig:a1a}
    \end{subfigure}
    \begin{subfigure}[t]{0.24\textwidth}
        \centering
        \includegraphics[width=\textwidth]{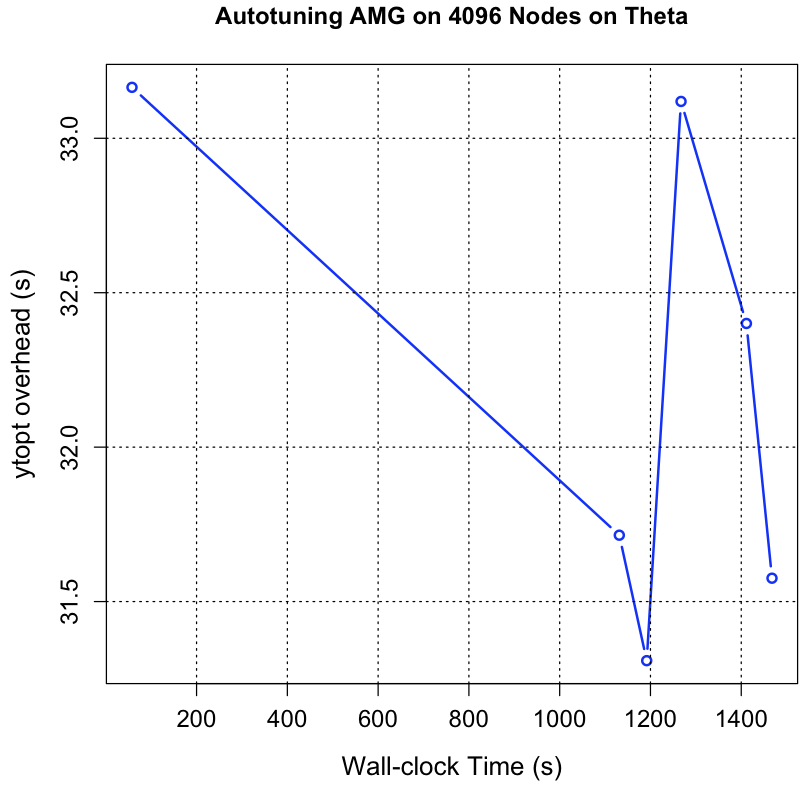}
        \subcaption{ytopt overhead}
        \label{fig:a1b}
    \end{subfigure}
    \setlength{\belowcaptionskip}{-8pt}
    \caption{Autotuning AMG at Large Scale on Theta}
    \label{fig:a1}
\end{figure}

Figure~\ref{fig:a1} shows the autotuning of AMG on 4,096 nodes on Theta. Because of the limited wall-clock time (1800 s), we see only six evaluations on 4,096 nodes, mainly caused by the second very long evaluation (1039.06 s) in Figure~\ref{fig:a1a}. We find the configuration for the long evaluation includes system parameters: 48 threads; OMP\_PLACES=threads (that are bound to specific logical processors); OMP\_PROC\_BIND= master (threads placed on master place to enhance locality); and OMP\_SCHEDULE=dynamic. We observe that the system parameter setting mainly causes the long application runtime because the first 48 cores of 64 cores are used and every two cores share the L2 cache. Figure~\ref{fig:a1b} still shows that the ytopt overhead is less than 34 s. 

\begin{figure}[ht]
    \centering
    \begin{subfigure}[t]{0.24\textwidth}
        \centering
        \includegraphics[width=\textwidth]{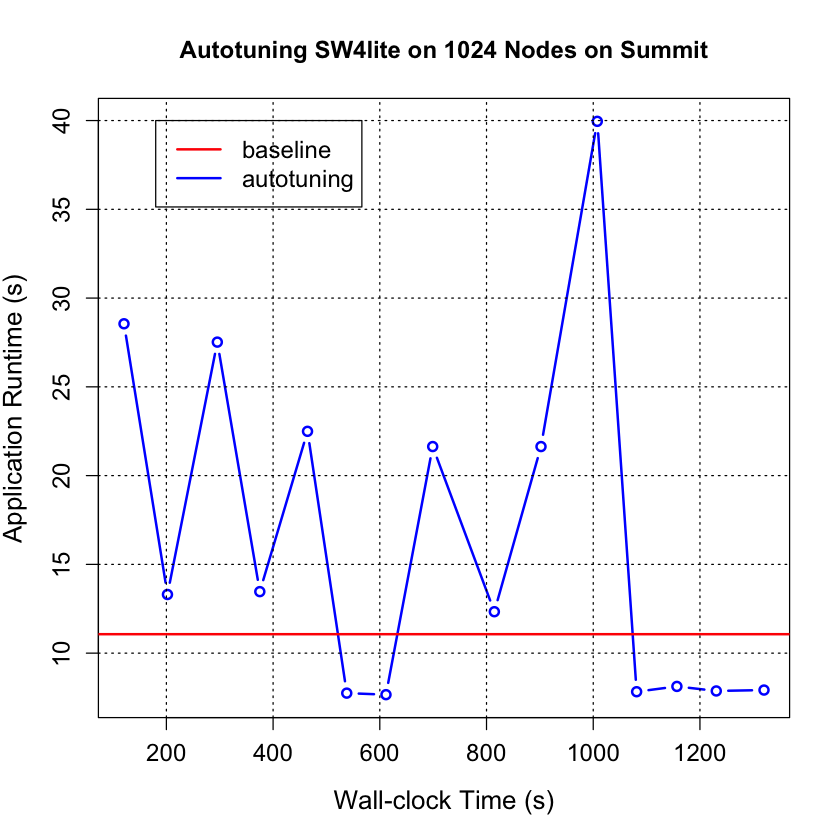}
        \subcaption{on 1024 nodes}
        \label{fig:w2a}
    \end{subfigure}
    \begin{subfigure}[t]{0.24\textwidth}
        \centering
        \includegraphics[width=\textwidth]{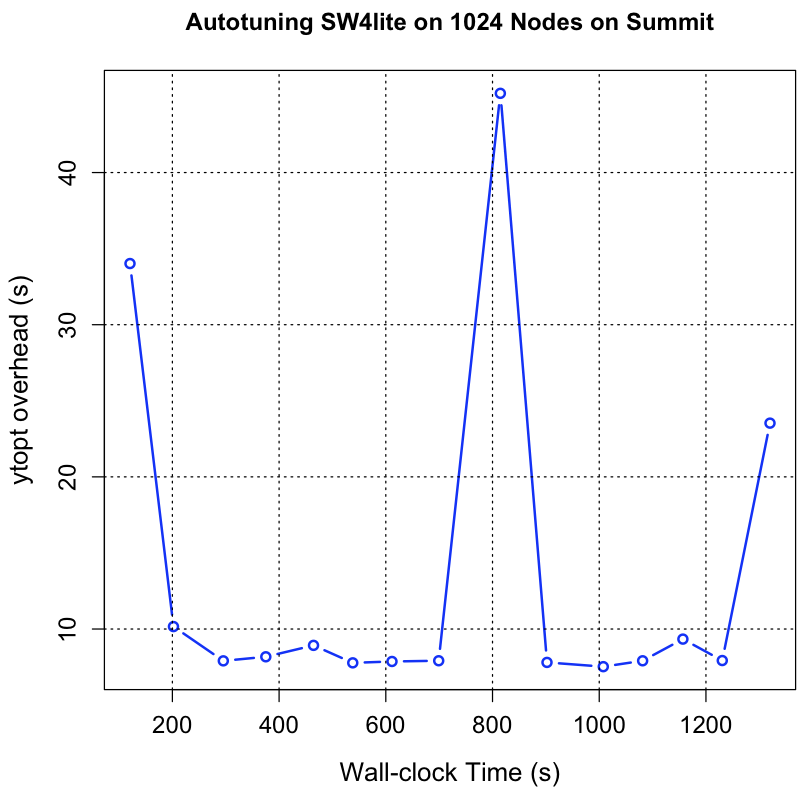}
        \subcaption{ytopt overhead}
        \label{fig:w2b}
    \end{subfigure}
    \setlength{\belowcaptionskip}{-8pt}
    \caption{Autotuning SW4lite at Large Scale on Summit}
    \label{fig:w2}
\end{figure}

SW4lite with the large problem LOH.1-h50 is strong scaling so that we can test it on up to 1,024 nodes. Figure~\ref{fig:w2} shows autotuning SW4lite on 1,024 nodes on Summit. As described in Table~\ref{tab:sz}, the parameter space size for this application is 2,211,840. As shown in Figure~\ref{fig:w2a}, at the beginning of the autotuning, ytopt samples the parameter space randomly for initial evaluations, then leverages the surrogate model to balance exploration of the search space and identifies more-promising parameter configurations using the LCB acquisition function. We observe that ytopt reaches the best configuration with the smallest runtime of 7.661 s, which is much better than  the baseline performance of 11.067 s. This is a 30.78\% performance improvement. Figure~\ref{fig:w2b} shows the ytopt overhead is less than 46 s.

\begin{figure}[ht]
    \centering
    \begin{subfigure}[t]{0.24\textwidth}
        \centering
        \includegraphics[width=\textwidth]{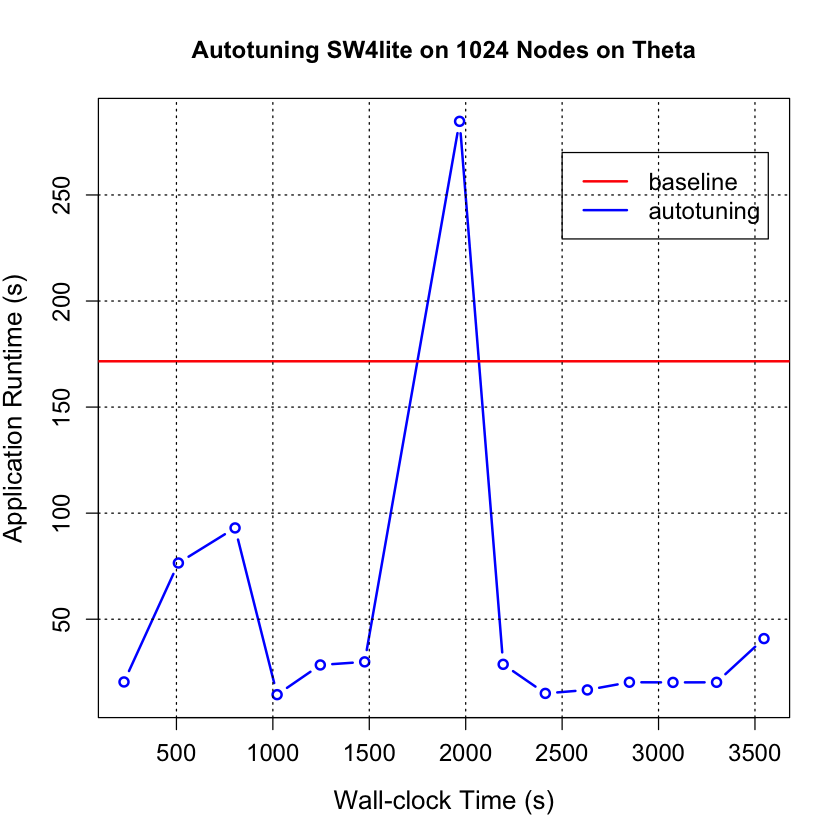}
        \subcaption{on 1024 nodes}
        \label{fig:w1a}
    \end{subfigure}
    \begin{subfigure}[t]{0.24\textwidth}
        \centering
        \includegraphics[width=\textwidth]{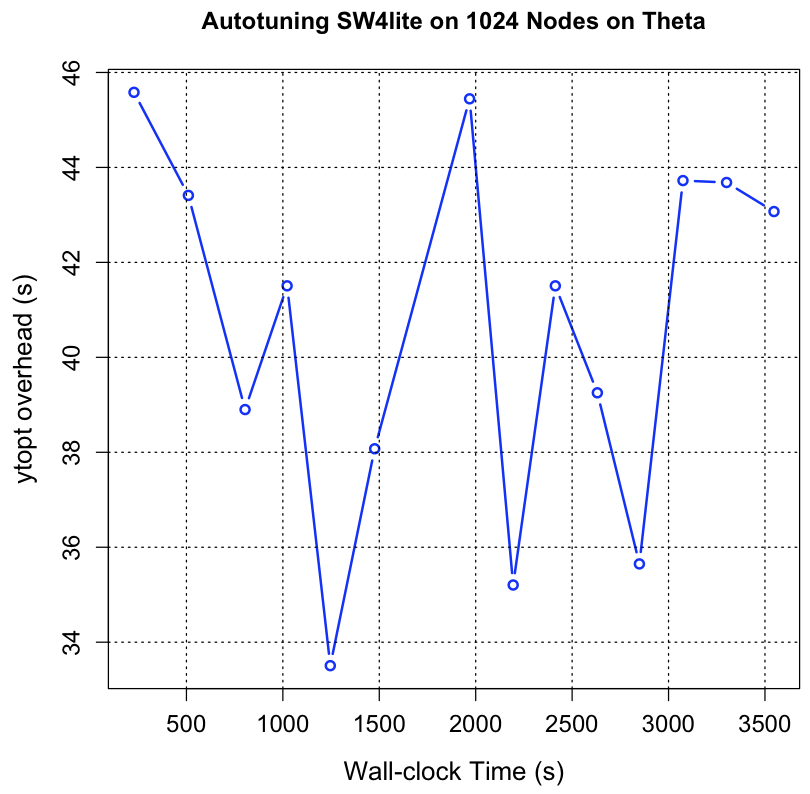}
        \subcaption{ytopt overhead}
        \label{fig:w1b}
    \end{subfigure}
    \setlength{\belowcaptionskip}{-8pt}
    \caption{Autotuning SW4lite at Large Scale on Theta}
    \label{fig:w1}
\end{figure}

Figure~\ref{fig:w1a} shows how SW4lite is autotuned on 1,024 nodes on Theta. We observe that ytopt reaches the best configuration with the smallest runtime of 14.427 s, which is much better than the baseline performance of 171.595 s. This is a 91.59\% performance improvement. We achieve the large improvement because we use the improved version of SW4lite \cite{WU21} to define the parameter space for SW4lite with the parameter MPI\_Barrier(MPI\_COMM\_WORLD). When running SW4lite on 1,024 nodes to measure the baseline performance, the compute time is small (around 3 s), but the communication time increases significantly (around 168 s) on Theta for the original code. Figure~\ref{fig:w1b} shows the ytopt overhead during the entire autotuning which is less than 46 s.

Overall, we observe that the ytopt overheads for the four applications are impacted mainly by the systems (for launching the application on the compute nodes) and application compiling times (given in Table~\ref{tab:cp}). %because the ytopt processing time includes the time spent in searching the parameter space, building the surrogate model, processing the new configuration to generate a new code and the aprun/jsrun command line, compiling the new code, launching the application, and storing the new configuration and performance in the performance database except the application runtime. 
We find that the ytopt overhead on up to 4,096 nodes on both Theta and Summit is less than 111 s \xingfu{in Table~\ref{tab:yo}}. 
This shows that our autotuning framework has low overhead and good scalability because the ytopt overhead does not increase much for autotuning the applications on small or large number of nodes. %In most cases, the autotuning framework performs faster on Summit than on Theta because Summit is much faster and more stable than Theta.

\begin{table}[ht]
\center
\caption{The maximum ytopt overhead (seconds) for each application on Theta and Summit}
\begin{tabular}{|r|c|c|c|c|c|}
\hline
System & XSBench-Mixed & XSBench & SWFFT & AMG & SW4lite  \\
\hline
Theta &  70 & 69 & 30  & 34   &   46\\
\hline
Summit & 24 & 111 & 50  & 45   &   46\\
\hline
\end{tabular}
\label{tab:yo}
\end{table}

\if 0
\begin{table}[ht]
\center
\caption{Performance improvement percentage (\%) for each application on Theta and Summit}
\begin{tabular}{|r|c|c|c|c|c|}
\hline
System & XSBench-mixed  & XSBench & SWFFT & AMG & SW4lite  \\
\hline
Theta & 1.6 & -9.8 & 0.03  & -8.21   &   91.59\\
\hline
Summit &2.8 &  -54.8  & 12.69  & 22.54   &   30.78\\
\hline
\end{tabular}
\label{tab:pi}
\end{table}

\fi

%% file: energy.tex
\section{Autotuning Energy at Large Scales}

In this section we apply the proposed energy autotuning framework in Figure~\ref{fig:en} to autotune the energy and EDP of four ECP proxy applications---XSBench, AMG, SWFFT, and SW4lite---on up to 4,096 nodes on Theta. Because energy consumption captures the tradeoff between the application runtime and power consumption and EDP captures the tradeoff between the application runtime and energy consumption, we use the autotuning framework to explore these tradeoffs for energy efficient application execution.

For measuring the baseline energy for each application with a given problem size, we set the number of threads to 64 on Theta and use GEOPM to run the application under the default system configuration five times. Then we use the smallest energy as the baseline for the application. 

For autotuning energy or EDP, after the evaluation of a configuration GEOPM generates the summary report gm.report, which records the package energy and DRAM energy for each node; we accumulate these as the node energy. When ytopt receives the report from GEOPM, it 
calculates an average node energy and uses that average energy as the primary metric for autotuning. Similarly, the average EDP is calculated.

\begin{figure}[ht]
    \centering
    \begin{subfigure}[t]{0.24\textwidth}
        \centering
        \includegraphics[width=\textwidth]{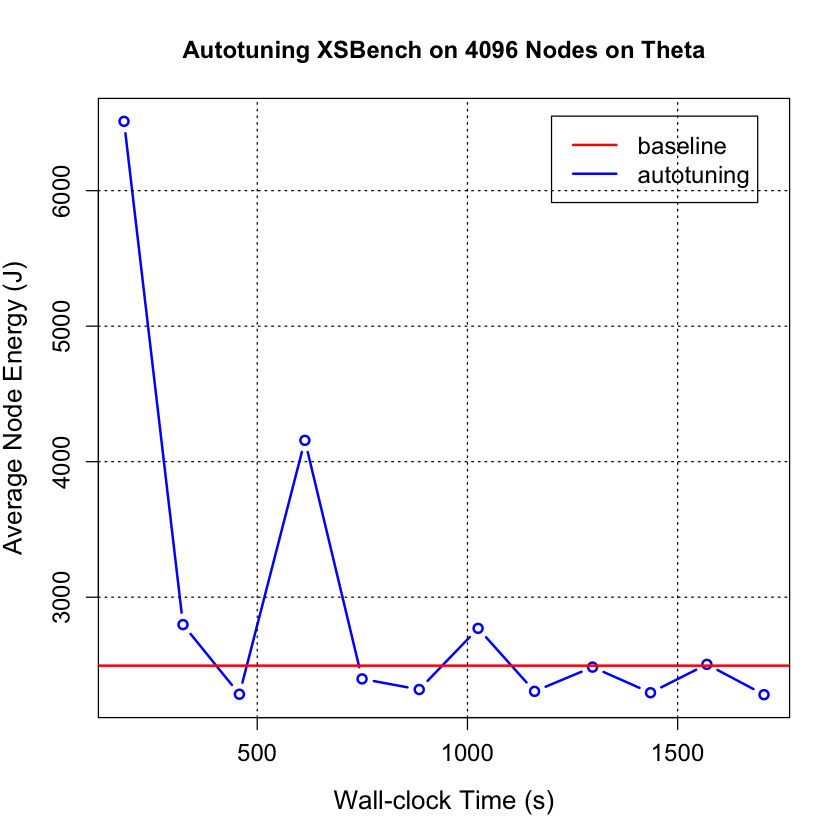}
        \subcaption{XSBench on 4096 nodes}
        \label{fig:x4}
    \end{subfigure}
    \begin{subfigure}[t]{0.24\textwidth}
        \centering
        \includegraphics[width=\textwidth]{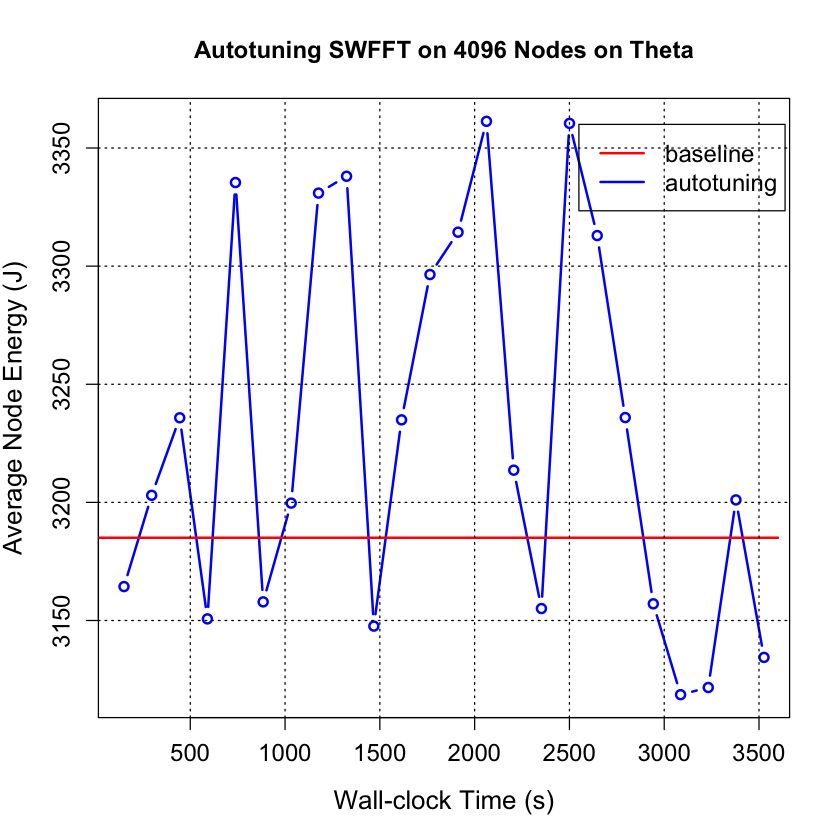}
        \subcaption{SWFFT on 4096 nodes}
        \label{fig:s3}
    \end{subfigure}
    \begin{subfigure}[t]{0.24\textwidth}
        \centering
        \includegraphics[width=\textwidth]{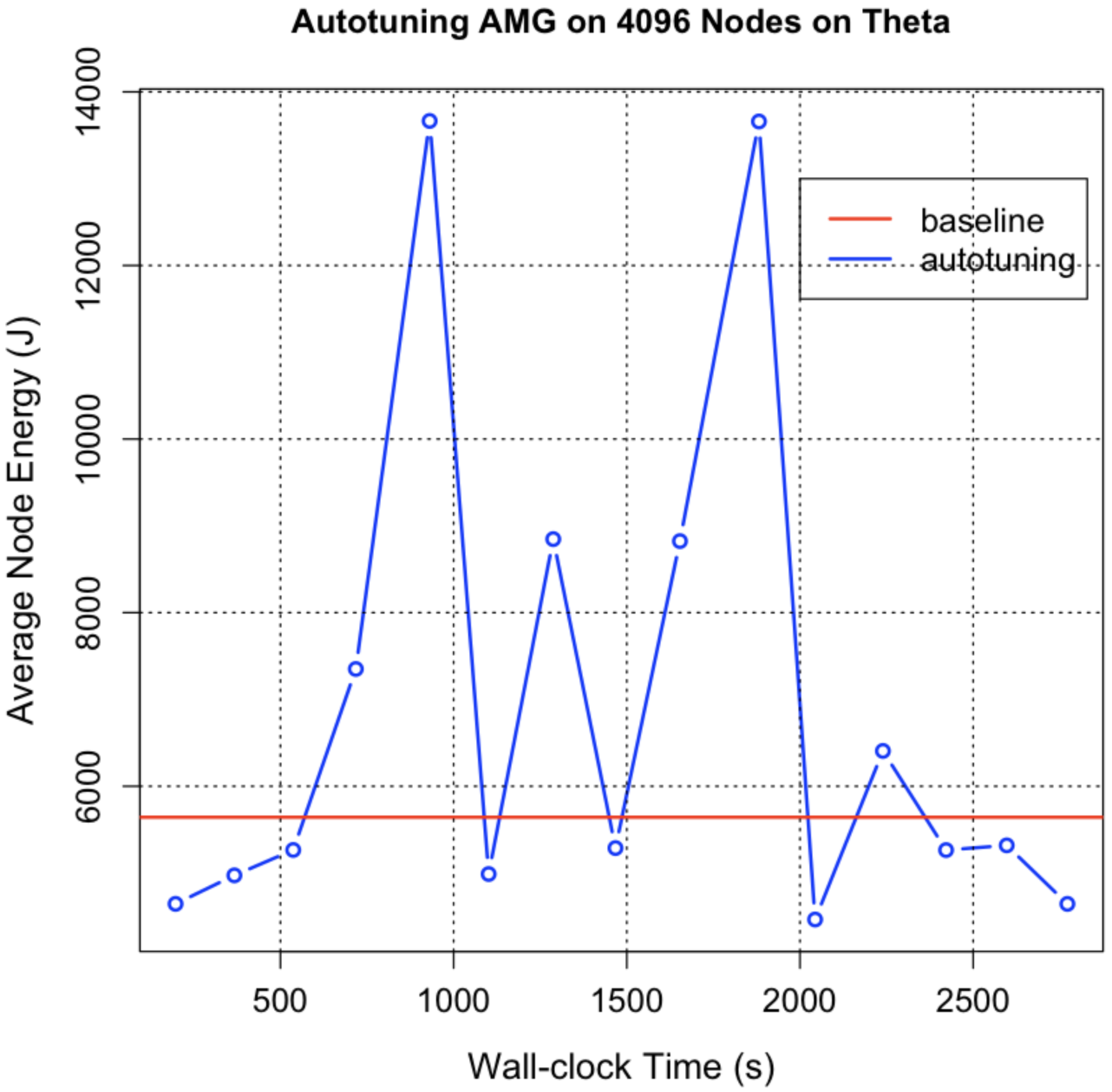}
        \subcaption{AMG on 4096 nodes}
        \label{fig:w3a}
    \end{subfigure}
    \begin{subfigure}[t]{0.24\textwidth}
        \centering
        \includegraphics[width=\textwidth]{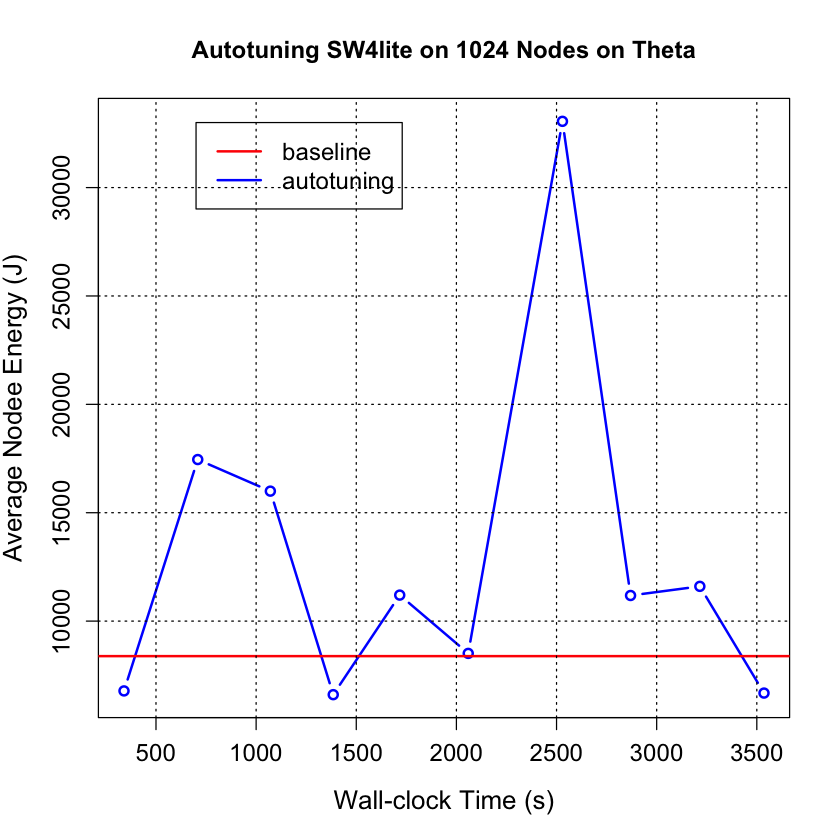}
        \subcaption{SW4lite on 1024 nodes}
        \label{fig:w3b}
    \end{subfigure}
    \setlength{\belowcaptionskip}{-8pt}
    \caption{Autotuning Energy at Large Scales on Theta}
    \label{fig:w3}
\end{figure}

Figure~\ref{fig:w3} shows using our energy framework to autotune the energy of the four ECP proxy applications at large scales on Theta. Figure~\ref{fig:x4} presents autotuning the energy of XSBench on 4,096 nodes, where the red line stands for the baseline node energy of 2494.905J. Using the framework achieves the lowest energy of 2280.806J. This is an 8.58\% energy savings. Figure~\ref{fig:s3} shows autotuning the energy of SWFFT on 4,096 nodes. The baseline node energy for SWFFT is 3185.027J.  Using the framework achieves the lowest node energy of 3118.604J. This is a 2.09\% energy savings. 

Figure~\ref{fig:w3a} presents autotuning the energy of AMG on 4,096 nodes. The baseline node energy for AMG is 5642.568J. Using the framework achieves the lowest node energy of 4566.747J. This is a 20.88\% energy saving. Figure~\ref{fig:w3b} presents autotuning the energy of SW4lite on 1,024 nodes. The baseline node energy for SW4lite is 8384.034J. Using the framework achieves the lowest node energy of 6606.233J. This is a 21.20\% energy saving. Compared with Figure~\ref{fig:w1a} for SW4lite, we identified the best configuration (32, 'sockets' ,'spread' ,'static', ' ', ' ' ,'\#pragma omp for nowait' ,' ' ) which resulted in 91.59\% performance improvement. The same configuration also resulted in the 21.20\% energy saving because the large performance improvement led to the energy saving. As we discussed before, the application runtime for SW4lite on 1024 nodes was dominated by the low power communication for the baseline, this was why the energy saving percentage is much less than the performance improvement percentage. Based on our observation, this is the case for other applications. 

\begin{figure}[ht]
    \centering
    \begin{subfigure}[t]{0.24\textwidth}
        \centering
        \includegraphics[width=\textwidth]{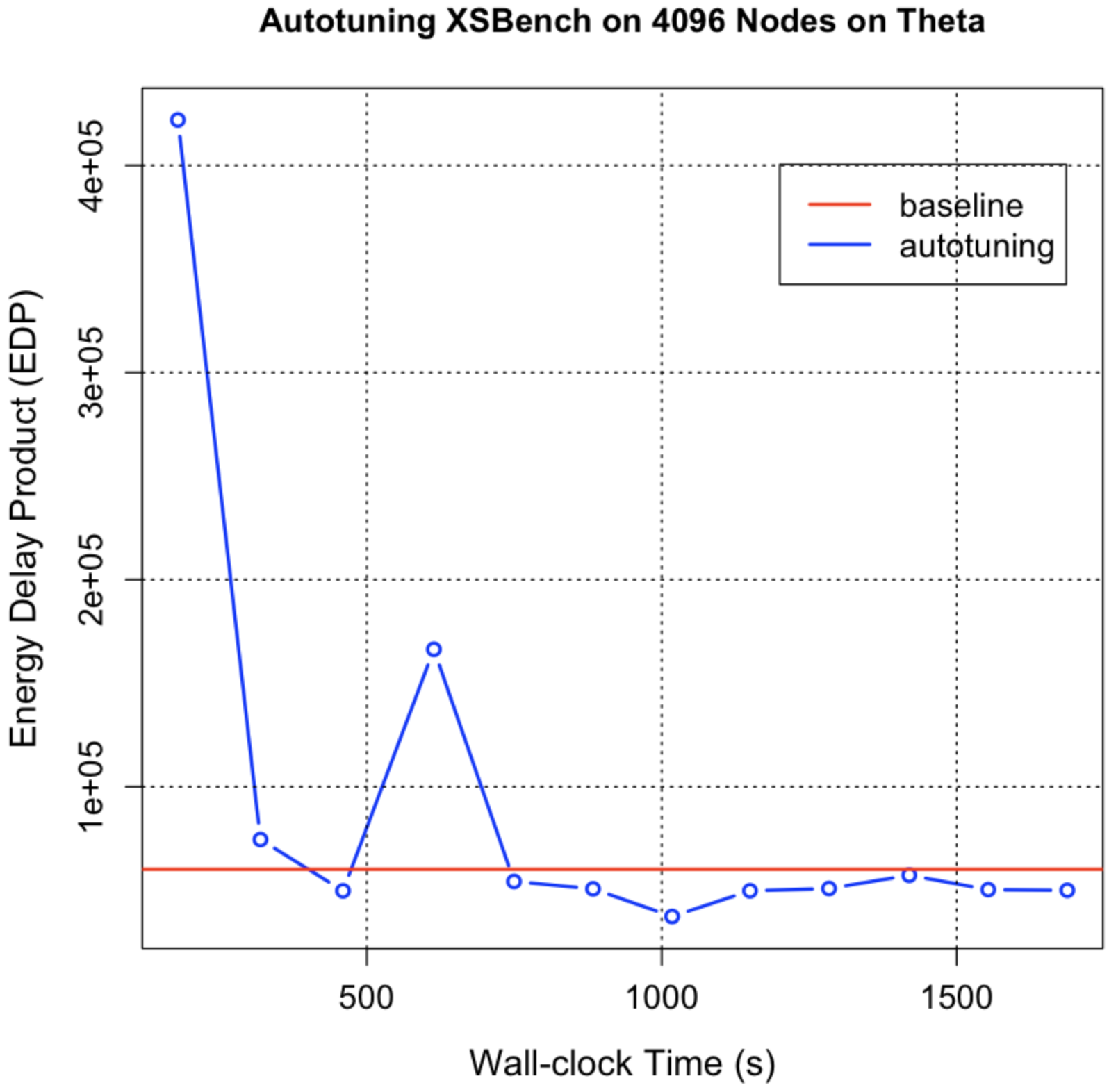}
        \subcaption{XSBench on 4096 nodes}
        \label{fig:x5}
    \end{subfigure}
    \begin{subfigure}[t]{0.24\textwidth}
        \centering
        \includegraphics[width=\textwidth]{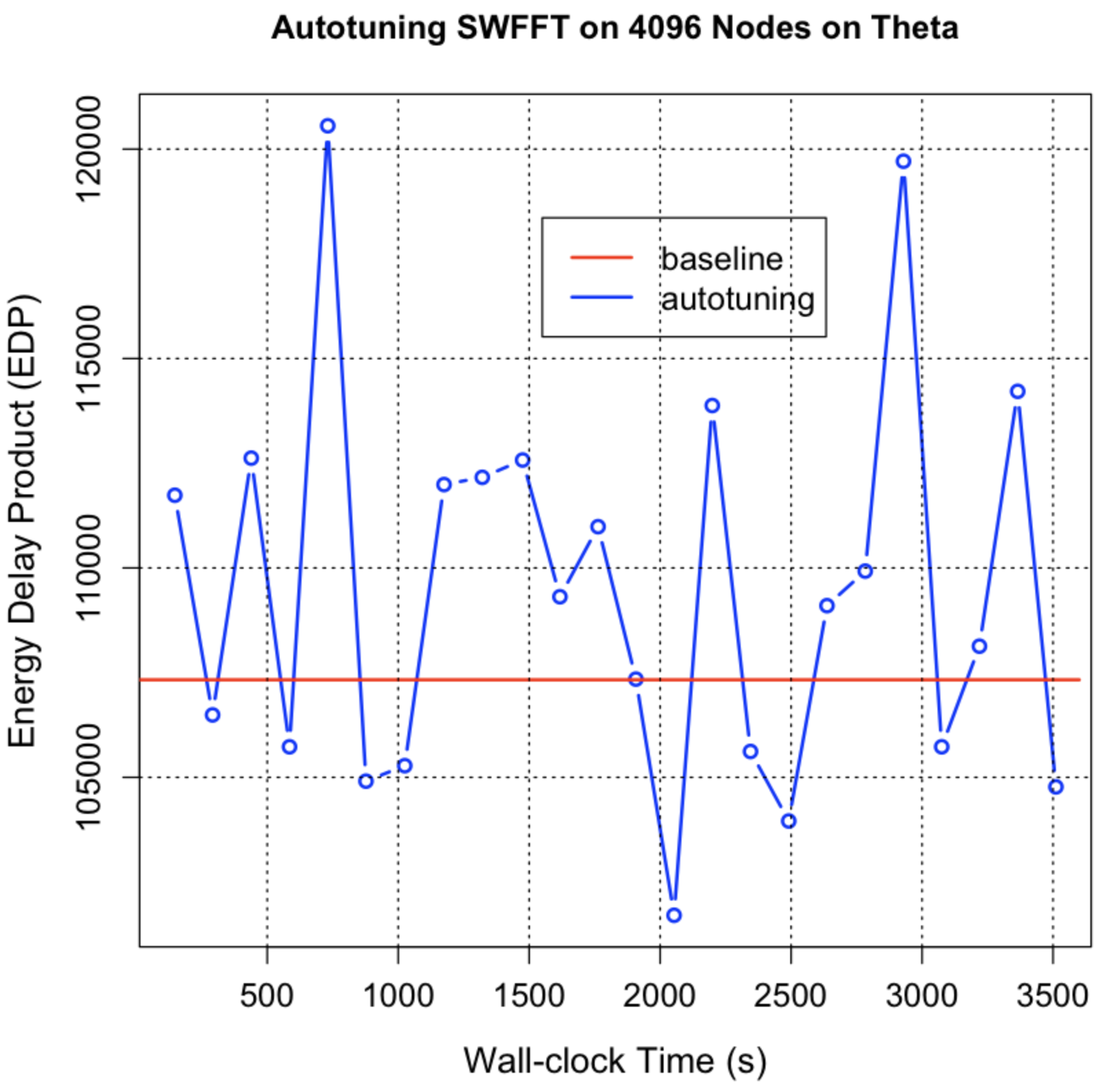}
        \subcaption{SWFFT on 4096 nodes}
        \label{fig:s4}
    \end{subfigure}
    \begin{subfigure}[t]{0.24\textwidth}
        \centering
        \includegraphics[width=\textwidth]{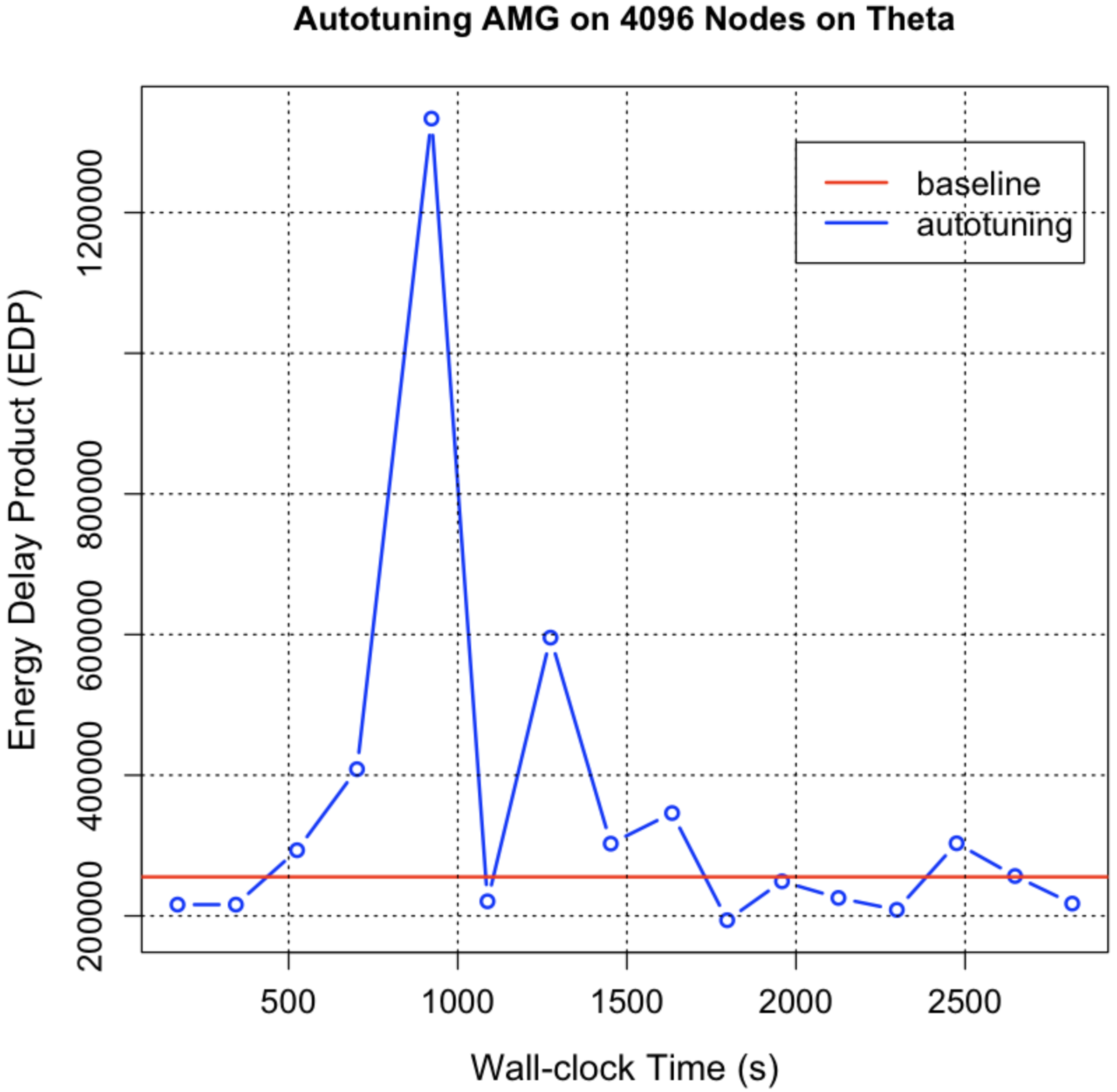}
        \subcaption{AMG on 4096 nodes}
        \label{fig:w4a}
    \end{subfigure}
    \begin{subfigure}[t]{0.24\textwidth}
        \centering
        \includegraphics[width=\textwidth]{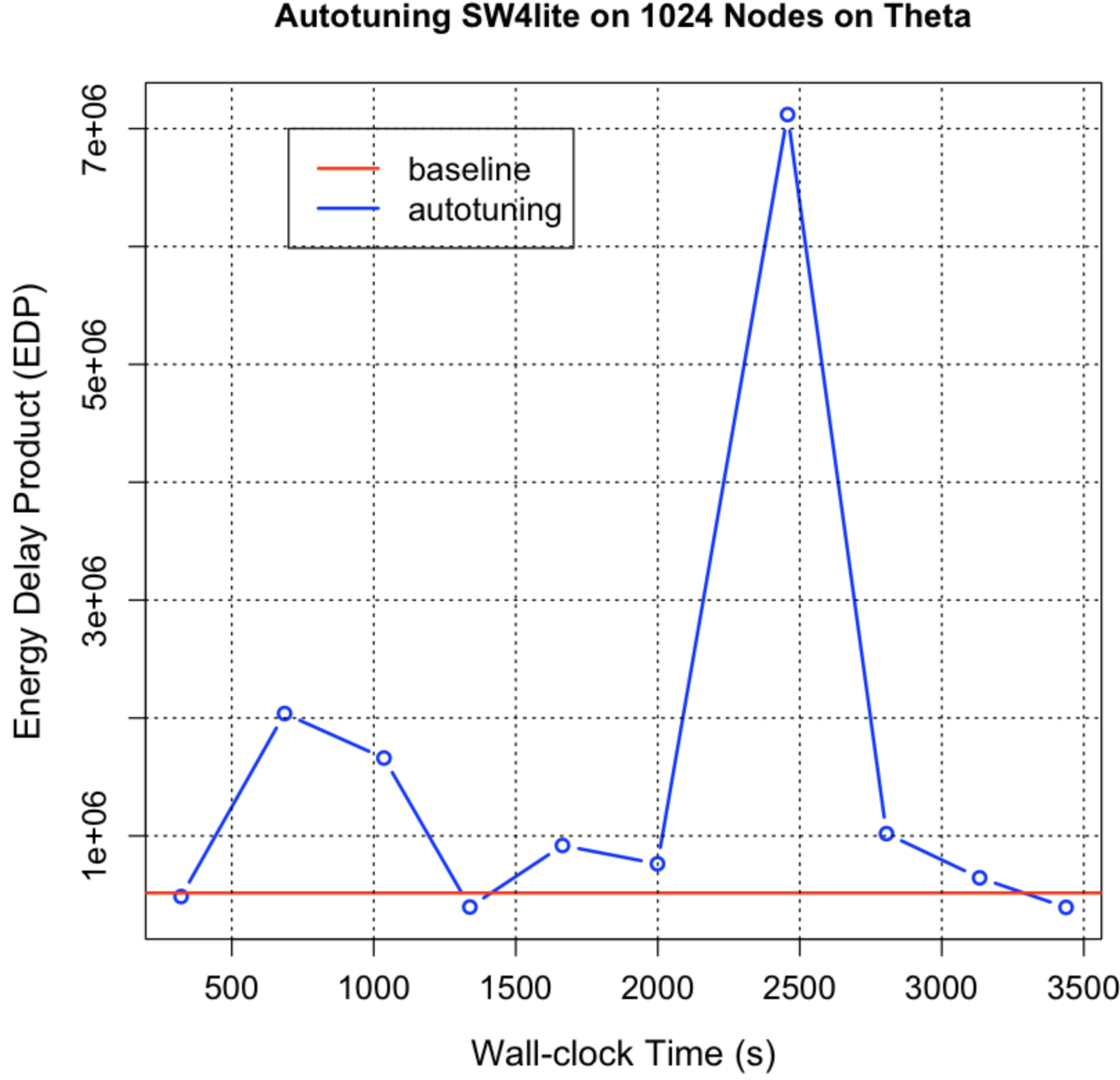}
        \subcaption{SW4lite on 1024 nodes}
        \label{fig:w4b}
    \end{subfigure}
    \setlength{\belowcaptionskip}{-8pt}
    \caption{Autotuning Energy Delay Product at Large Scales on Theta}
    \label{fig:w4}
\end{figure}

Figure~\ref{fig:w4} shows using our energy framework to autotune the EDP of the four ECP proxy applications at large scales on Theta. Figure~\ref{fig:x5} presents autotuning the energy of XSBench on 4,096 nodes, where the red line stands for the baseline node EDP. Using the framework achieves the lowest EDP with 37.84\% improvement. Figure~\ref{fig:s4} shows autotuning the energy of SWFFT on 4,096 nodes. Using the framework achieves the lowest EDP with 5.24\% improvement.
Figure~\ref{fig:w4a} presents autotuning the energy of AMG on 4,096 nodes. Using the framework achieves the lowest EDP with 24.13\% improvement. Figure~\ref{fig:w4b} presents autotuning the energy of SW4lite on 1,024 nodes. Using the framework achieves the lowest EDP with 23.70\% improvement.
Because EDP is the product of energy and application runtime, the EDP improvement is better than the energy improvement shown in Table \ref{tab:es}. The best configuration for using EDP as the metric is similar to that for using energy as metric.

Overall, using our energy autotuning framework to identify the best configurations for the four ECP proxy applications results in up to 21.2\% energy savings and up to 37.84\% improvement in EDP on up to 4,096 nodes shown in Table \ref{tab:es}. This aids us in exploring the tradeoffs between application runtime and power/energy for energy efficient application execution.

\begin{table}[ht]
\center
\caption{Improvement percentage (\%) for each application on Theta}
\begin{tabular}{|r|c|c|c|c|}
\hline
Theta & XSBench & SWFFT & AMG & SW4lite  \\
\hline
Energy &  8.58 & 2.09  & 20.88   &   21.20\\
\hline
EDP &  37.84 & 5.24  & 24.13   &   23.70\\
\hline
\end{tabular}
\label{tab:es}
\end{table}

%% file: conclusions.tex
\section{Conclusions}

%Efficiently utilizing power and optimizing the performance of scientific applications under power and energy constraints are challenging tasks in scientific applications. To address these issues, 

In this paper, we proposed the low-overhead autotuning frameworks to autotune four hybrid MPI/OpenMP ECP proxy applications---XSBench, AMG, SWFFT, and SW4lite---at large scales and explored the tradeoffs between application runtime and power/energy for energy efficient application execution. We used Bayesian optimization with a Random Forest surrogate model to effectively search the parameter spaces with up to 6 million different configurations on Theta and Summit. We used the autotuning framework to explore the tradeoffs between application runtime and power/energy for energy efficient application execution. The experimental results showed that our autotuning framework had low overhead and good scalability. By using the autotuning framework to identify the best configuration, we achieved up to 91.59\% performance improvement, up to 21.2\% energy savings, and up to 37.84\% EDP improvement on up to 4,096 nodes. The ytopt autotuning framework is open source and available to download from the link in \cite{YTO}. 

For future work, we will improve the framework overhead by reducing the application compiling time with pre-compiling the unchanged code files and setting a proper evaluation timeout to evaluate more good configurations.
Our current autotuning framework uses Ray \cite{RAY} to do one evaluation each time; this affected the effectiveness of identifying the promising search regions at the beginning of the autotuning. We plan to extend the framework to do multiple evaluations in parallel using libensemble \cite{libem} to improve the initial effectiveness. 
We also plan to add transfer learning \xingfu{and online tuning} to the framework so that it can transfer what it learns from the applications at a small scale in problem sizes and system sizes to guide and/or predict \xingfu{the best configurations for} autotuning at large scales.